\documentclass{article} 
\usepackage{iclr2027_conference,times}

\usepackage{amsmath,amsfonts,bm}

\def\eqref#1{Eq.~\ref{#1}}

\def\1{\bm{1}}

\DeclareMathAlphabet{\mathsfit}{\encodingdefault}{\sfdefault}{m}{sl}
\SetMathAlphabet{\mathsfit}{bold}{\encodingdefault}{\sfdefault}{bx}{n}

\usepackage{hyperref}
\usepackage{url}

\usepackage{amssymb}
\usepackage{algorithm}
\usepackage{algpseudocode}
\usepackage{array}
\usepackage{booktabs}
\usepackage{colortbl}
\usepackage{enumitem}
\usepackage{float}
\usepackage{graphicx}
\usepackage{longtable}
\usepackage{makecell}
\usepackage{multirow}
\usepackage{pifont}
\usepackage{subcaption}
\usepackage{tabularx}
\usepackage{wrapfig}
\usepackage{xcolor}

\newcolumntype{L}[1]{>{\raggedright\arraybackslash}p{#1}}
\newcolumntype{Y}{>{\raggedright\arraybackslash}X}

\title{POEM: Output-Relative Block-Sparse \\ Attention for Long-Context Prefilling}

\author{
Haoyi Wu$^{1,*}$ \quad
Yunkai Dang$^{1,*}$ \quad
Wanyu Zhang$^{1}$ \quad
Lei Wang$^{3}$ \quad
Wenbin Li$^{1,2,\dagger}$ \\[0.8em]
$^{1}$State Key Laboratory of Novel Software Technology,
Nanjing University, Nanjing, China \\
$^{2}$Shenzhen Research Institute of Nanjing University,
Shenzhen, China \\
$^{3}$University of Wollongong, Australia
}

\iclrfinalcopy 
\begin{document}

\maketitle \fancyhead[L]{}

\begingroup
\renewcommand{\thefootnote}{\fnsymbol{footnote}}

\footnotetext[1]{$^{*}$Equal contribution.}

\footnotetext[2]{$^{\dagger}$Corresponding author:
\texttt{liwenbin@nju.edu.cn}.}

\endgroup

\begin{abstract}
As large language models (LLMs) evolve to support increasingly long contexts, accelerating prefilling without compromising output accuracy becomes a critical challenge.
Block-sparse attention addresses this by retaining only the most important key-value blocks, but existing importance proxies often discard blocks and cause output errors.
These proxies generally estimate a block's contribution to the dense attention output (i.e., attention-weighted value), implicitly assuming that removing the block loses exactly this contribution.
In this work, we show that this assumption does not hold because attention renormalization amplifies the retained blocks' contributions.
As a result, the output error depends not only on the removed attention mass but also on the distance between the removed block's output and the dense output.
We call the distance \textit{output deviation} and empirically find that it varies by over 20${\times}$ even among blocks with comparable attention mass, so existing proxies often discard blocks whose large output deviation leads to large output errors.
To address this gap, we propose \textit{Prefilling via Output Error Minimization (POEM)}, a training-free method that combines each block's contribution and output deviation into an output-relative score, reducing the output error of each head.
To connect the head-level optimization to the model's predictions, we derive a bound on the prediction divergence that weights each head's output error by its sensitivity, which guides POEM to calibrate the retention threshold of each head.
On long-context benchmarks, POEM maintains accuracy parity with dense attention while achieving up to a $6.5{\times}$ attention speedup over FlashAttention-2.
\end{abstract}

\begingroup
\raggedbottom
\setlength{\parskip}{4pt}
\setlength{\intextsep}{6pt}
\setlength{\textfloatsep}{8pt}

\section{Introduction}
\label{sec:introduction}
Large language models (LLMs) \citep{brown2020language,llama2023llama} have achieved striking advances in recent years, demonstrating strong performance across a wide range of tasks. These advances have enabled their deployment in increasingly ambitious applications, from scientific discovery \citep{novikov2025alphaevolve,yamada2025ai} to autonomous agents \citep{wang2023voyager,zhang2026darwin}.
Such applications often demand processing information across extremely long sequences such as extensive scientific corpora and long-horizon task trajectories \citep{bai2025longbench}.
However, as context lengths continue to grow, the quadratic cost of self-attention \citep{vaswani2023attentionneed} makes prefilling prohibitively expensive, rendering attention the dominant computational bottleneck in LLM inference \citep{fu2024challenges}.
To alleviate this bottleneck, recent studies have developed efficient attention mechanisms \citep{yuan2025native,hatamizadeh2026gated,sun2025efficient}, ranging from sparsifying token interactions to reformulating attention with subquadratic alternatives.

Among these mechanisms, block-sparse attention is particularly attractive, as it retains exact attention over the relevant regions while entirely skipping the rest.
Block-sparse attention \citep{child2019generating} operates by partitioning the key--value sequence into fixed-size blocks, retaining only important blocks for each query, and computing exact attention over the retained subset.
A key challenge is to accurately identify the most important blocks without computing the full attention matrix \citep{roy-etal-2021-efficient}.
Recent works on block-sparse attention have explored a wide range of strategies.
Some methods incorporate block selection into model training \citep{yuan2025native,lu2026moba} or train auxiliary gates to predict block importance \citep{gao2024seerattention}.
However, these methods incur additional training costs, complicating their application to existing pretrained LLMs.
To avoid this overhead, training-free methods instead estimate block importance using lightweight proxies~\citep{lai2025flexprefill}.
These methods select blocks by measuring their contribution to dense attention.
Most of them estimate attention mass \citep{jiang2024minference,lai2025flexprefill,xu2025xattention,wang2026prismspectralawareblocksparseattention}, while STEM \citep{niu2026stem} further incorporates value to account for differences in value contributions.
Nevertheless, a block's contribution to dense attention is not equivalent to the output shift by its removal.
When a block is dropped, softmax renormalization amplifies the contributions of the retained blocks, compensating for part of the removed blocks' contribution.

So what determines whether a removed block's contribution is easily compensated?
As Figure~\ref{fig:intro_output_relative} (left) illustrates, if a block's output lies close to the dense output, renormalization amplifies the retained blocks' contributions while largely preserving the output direction, resulting in only a small output shift.
However, if the block's output points away from the dense output, the same amplification cannot compensate for its removal, and the output shifts substantially.
We call the distance between a block's output and the dense output the block's \emph{output deviation}.
Empirically, we observe that output deviation varies by more than $20{\times}$ across blocks of comparable attention mass, forming distinct low-deviation and high-deviation modes (center).
Therefore, existing proxies that rely on block contribution alone cannot distinguish safely removable blocks from critical ones.

\begin{figure}[!t]
  \centering
  \includegraphics[width=0.9\linewidth]{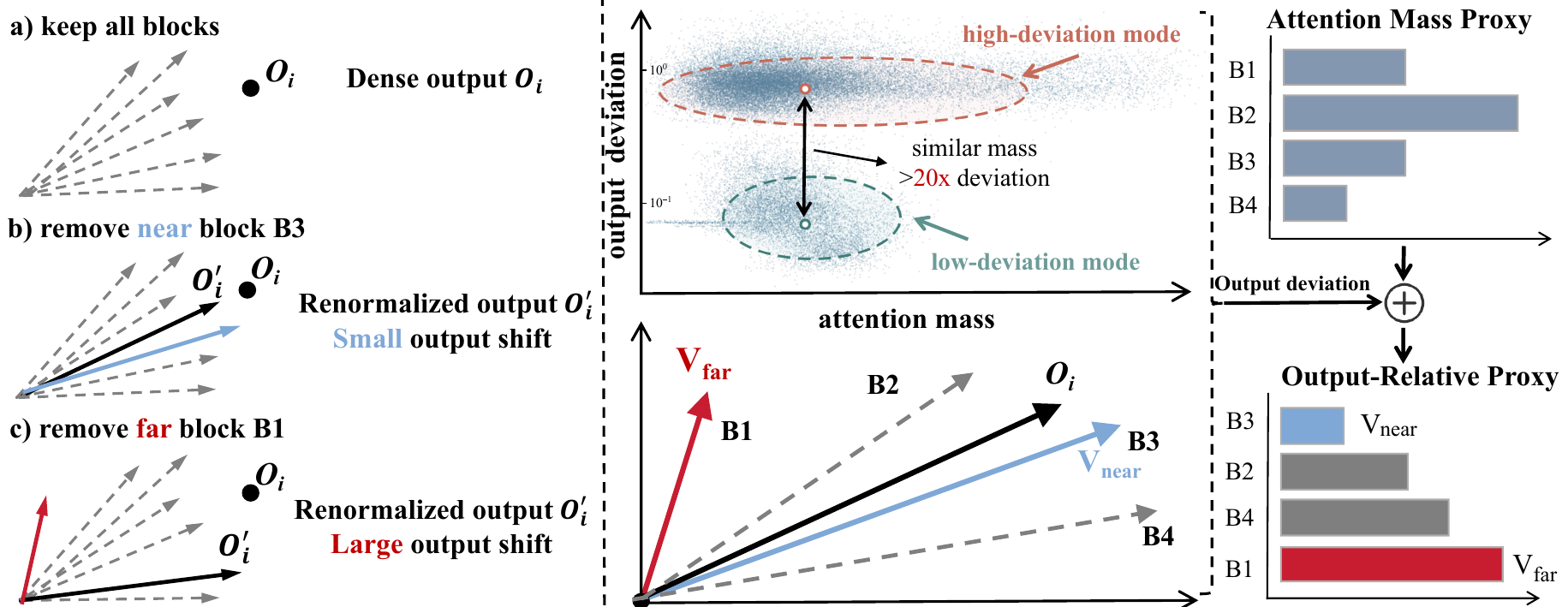}
  \captionsetup{font=normalsize,skip=4pt,justification=justified,singlelinecheck=false}
  \caption{Through experiments, we observe that blocks with similar attention mass can differ by over $20{\times}$ in output deviation.
Removing a high-deviation block (B1) shifts the attention output far more than removing a low-deviation one (B3) of comparable mass.
We therefore combine attention mass with output deviation into our output-relative proxy.
V denotes the exact output of a block.}
  \label{fig:intro_output_relative}
\end{figure}

To address this gap, we propose \textit{Prefilling via Output Error Minimization (POEM)}, a training-free block-sparse attention method.
\textbf{(1) Output-relative block selection:} Motivated by the renormalization analysis above, we derive a head-level output-relative score that combines each block's attention mass with the deviation of its output from the dense head output.
A two-stage pipeline makes this score practical: mass estimates identify a candidate pool at low cost, and the output-relative score then selects the final blocks within it.
To connect this head-level optimization to model predictions, we derive a theoretical bound on predictive divergence that weights each head's output error by its sensitivity.
This bound motivates \textbf{(2) Offline threshold calibration:} POEM calibrates a separate retention threshold for each head group according to its sensitivity.
Because calibrating every head group independently would be prohibitively expensive, POEM ranks heads by sensitivity and jointly calibrates similarly ranked heads via an efficient batched bisection procedure.

Experiments on LongBench~\citep{bai2024longbench} and RULER~\citep{hsieh2024ruler} using Llama-3.1-8B~\citep{grattafiori2024llama}, Qwen3-8B, and Qwen3-14B~\citep{yang2025qwen3} show that POEM maintains accuracy comparable to that of full attention at high sparsity.
On RULER at 128K context, POEM achieves a 2.4-point gain over our mass-only baseline, demonstrating the effectiveness of the output-relative score.
POEM also achieves the lowest attention latency among the compared methods, achieving up to a $6.5\times$ speedup over FlashAttention-2~\citep{dao2024flashattention} at a 128K context.

Our contributions can be summarized as:
\begin{itemize}[leftmargin=*,topsep=3pt,itemsep=2pt,parsep=0pt,partopsep=0pt]
\item We propose POEM, a training-free block-sparse attention method that combines attention mass with output deviation for output-relative block selection, optimizing output error within heads.

\item To bridge head-level optimization and model-level prediction, we derive a bound on output divergence in terms of head error weighted by sensitivity, motivating our offline threshold calibration.

\item Across three LLMs on LongBench and RULER, POEM almost matches full-attention accuracy at high sparsity and achieves up to a $6.5{\times}$ attention speedup at 128K prefill.

\end{itemize}
\par
\endgroup

\section{Related Work}
\label{sec:related_work}

\textbf{Training-free sparse attention.}
Training-free methods use the sparsity already present in pretrained attention.
Prefilling must select blocks for every query position, so the selector itself has to be cheap yet accurate.
To score blocks, SALE~\citep{ji2025sale} uses low-bit estimates, Sketch\&Walk~\citep{le2026scout} propagates randomized sketches across layers, UniSparse~\citep{liu2025unified} compresses and routes at multiple granularities, and Prism~\citep{wang2026prismspectralawareblocksparseattention} keeps high-frequency positional signals with spectral-aware scores.
Other work lowers cost elsewhere: SpargeAttention~\citep{zhang2025spargeattention} co-designs the selector with the kernel, Pbs-Attn~\citep{wang2026sparserblocksparseattentiontoken} packs important tokens into fewer blocks, and ProxyAttn~\citep{wang2026proxyattnguidedsparseattention} lets a few representative heads score for the rest.
These selectors rank blocks mainly by attention mass, and STEM~\citep{niu2026stem} adds value norms to estimate how much each block contributes to dense attention.
POEM instead aims to minimize the output error caused by removal, combining attention mass and output deviation together to select the blocks.

\textbf{Training-based sparse attention.}
Training-based methods learn sparsity patterns during pretraining or adaptation.
More recent designs include Native Sparse Attention~\citep{yuan2025native}, which combines compression, selection, and local attention; MoBA~\citep{lu2026moba}, which routes each query to its top-$k$ blocks.
InfLLM-V2~\citep{zhao2026infllm}, which switches between dense and sparse attention depending on sequence length, and HiLS~\citep{hu2026hierarchical}, which learns hierarchical retrieval from the language-modeling loss.
SeerAttention~\citep{gao2024seerattention} instead trains block-level gates on a frozen backbone for prefilling, and SeerAttention-R~\citep{gao2025seerattention} extends them to decoding.
The selectors also differ in supervision: DeepSeek Sparse Attention~\citep{liu2025deepseek} uses a token-level indexer with attention scores, whereas SAS~\citep{li2026sas} ranks context with the language-modeling loss.
All of these methods train either the model or an auxiliary selector, while POEM is training-free and applies directly to pretrained models at no additional training cost.

\section{Preliminaries}
\label{sec:preliminaries}

\noindent\textbf{Attention and block-sparse formulation.}
Let $\mathbf Q,\mathbf K,\mathbf V\in\mathbb R^{N\times d_h}$ be the query, key, and value matrices with row vectors $\mathbf q$, $\mathbf k$, $\mathbf v$, where $N$ is the sequence length and $d_h$ the head dimension. Under causal masking, the attention weight and output are defined for each query position as follows:
\begin{equation}
A_{it}
=
\frac{\exp(\mathbf q_i\mathbf k_t^\top/\sqrt{d_h})}
{\sum_{u=1}^{i}\exp(\mathbf q_i\mathbf k_u^\top/\sqrt{d_h})},
\qquad
\mathbf o_i=\sum_{t=1}^{i}A_{it}\mathbf v_t .
\label{eq:dense_attention_probability}
\end{equation}
Block-sparse attention~\citep{xu2025xattention} partitions KV sequences into $B$ blocks $\{\mathcal B_b\}_{b=1}^{B}$ and restricts each query $i$ to a retained block set $\mathcal S_i\subseteq\{1,\dots,B\}$. For block $b$, define the block attention mass $m_{ib}=\sum_{t\in\mathcal B_b}A_{it}$ and the block output $\bar{\mathbf v}_{ib}=m_{ib}^{-1}\sum_{t\in\mathcal B_b}A_{it}\mathbf v_t$, so that $\mathbf o_i=\sum_b m_{ib}\bar{\mathbf v}_{ib}$. Given the discarded mass $\delta_i=\sum_{b\notin\mathcal S_i}m_{ib}$, the renormalised sparse output and the projected output of a layer $\ell$ with $H$ heads are respectively given by the following expressions:
\begin{equation}
\widetilde{\mathbf o}_i
=
\frac{\sum_{b\in\mathcal S_i}m_{ib}\bar{\mathbf v}_{ib}}
{1-\delta_i},
\qquad
\mathbf o_{\ell}^{\mathrm{out}}=\sum_{h=1}^{H}\mathbf o_{\ell,h}\,\mathbf W_{O,\ell}^{(h)},
\label{eq:sparse_attention_output}
\end{equation}
where ${\mathbf W_{O,\ell}^{(h)}\in\mathbb R^{d_h\times d}}$ projects the output $\mathbf o_{\ell,h}$ of head $h$ to the model dimension $d$.

\section{Method}
\label{sec:method}
\begin{figure}[!t]
  \centering
  \includegraphics[width=0.9\textwidth]{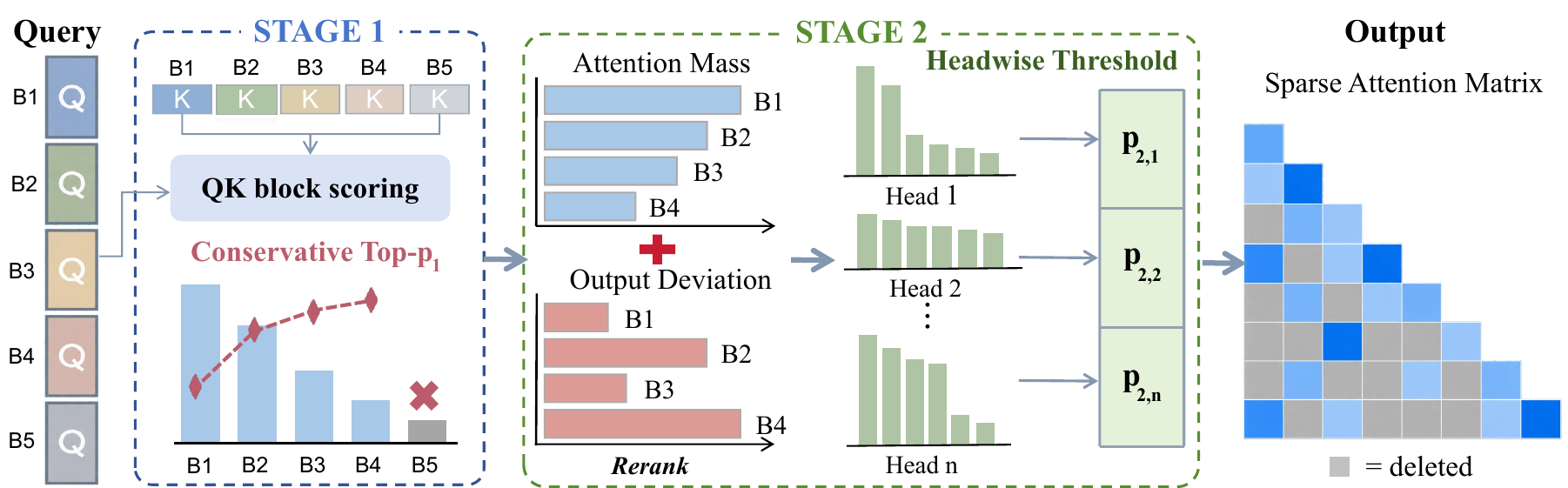}
  \caption{Overview of POEM. Stage~1 preselects candidate blocks using
  coarse attention mass.
  Stage~2 reranks candidates with output-relative scores and selects blocks
  using calibrated headwise thresholds.
  $p_1$ is the Stage~1 threshold, and $p_{2,h}$ is the calibrated threshold for head $h$.}
  \label{fig:poem_framework}
\end{figure}

In this section, we first present the output-relative block score (Section~\ref{sec:output_relative_block_selection}).
We then introduce a hierarchical error decomposition (Section~\ref{sec:hierarchical_error}) that connects predictive divergence to head output error, followed by the offline headwise
threshold calibration (Section~\ref{sec:headwise_budget}).
Finally, we summarize the resulting POEM pipeline with two stages
(Section~\ref{sec:two_stage_pipeline}) and show it in Figure~\ref{fig:poem_framework}.
\subsection{Output-relative block selection}
\label{sec:output_relative_block_selection}
\begin{wraptable}{R}{0.45\textwidth}
  \centering
  \caption{Comparison of proxies.
  $\ast$ means antidiagonal scoring, and $v^{\max}$ denotes the maximum value norm within the block.}
  \label{tab:proxy_comparison}
  \resizebox{\linewidth}{!}{%
    \begin{tabular}{@{}lccc@{}}
      \toprule
      Method & Proxy Score & \makecell{Value-\\Aware}
        & \makecell{Output-\\Relative} \\
      \midrule
      MInference
        & $\displaystyle
            \frac{\mathbf{q}\mathbf{k}^{\top}}{\sqrt{d}}$
        & \ding{55} & \ding{55} \\
      FlexPrefill
        & $\displaystyle
            \frac{\mathbf{q}\mathbf{k}^{\top}}{\sqrt{d}}$
        & \ding{55} & \ding{55} \\
      XAttention
        & $\displaystyle
            \frac{\mathbf{q}\mathbf{k}^{\top}}{\sqrt{d}}{}^{\ast}$
        & \ding{55} & \ding{55} \\
      STEM
        & $\displaystyle
            \frac{\mathbf{q}\mathbf{k}^{\top}}{\sqrt{d}}{}^{\ast}
            +\beta\operatorname{ReLU}\!\left(\log v^{\max}\right)$
        & \ding{51} & \ding{55} \\
      \addlinespace[4pt]
      \textbf{POEM}
        & $\displaystyle
            \frac{\mathbf{q}\mathbf{k}^{\top}}{\sqrt{d}}{}^{\ast}
            +\beta\log\left\|
              \mathbf{v}
              +\frac{\mathbf{q}}{\sqrt{d}}\mathbf{C}
              -\widehat{\mathbf{o}}
            \right\|$
        & \ding{51} & \ding{51} \\
      \bottomrule
    \end{tabular}%
  }
\end{wraptable}
A central design question for training-free block-sparse attention is how to reliably select significant blocks under a fixed budget while preserving the dense output.
Existing methods such as MInference~\citep{jiang2024minference}, FlexPrefill~\citep{lai2025flexprefill}, and XAttention~\citep{xu2025xattention} score each block solely by the attention mass.
STEM~\citep{niu2026stem} furthermore augments the QK score with the largest value norm inside the block, which makes the proxy value-aware.
However, all of existing methods overlook the effect of renormalization, leaving out output deviation, except POEM.
POEM's proxy scores each block by combining attention mass with output deviation, making the proxy both value-aware and output-relative, as summarized in Table~\ref{tab:proxy_comparison}.
This subsection then introduces POEM's output-relative proxy and describes how it is ultimately used in the complete POEM pipeline.

\textbf{From output error to block importance.} Our objective is to select a subset of blocks $\mathcal S_i$ that minimizes the output error between the sparse output $\widetilde{\mathbf{o}}_i$ and the dense output $\mathbf{o}_i$.
As defined in Section~\ref{sec:preliminaries}, we express this objective as the following subset selection problem:
\begin{equation}
    \min_{\mathcal S_i}\left\|\mathbf{o}_i-\widetilde{\mathbf{o}}_i\right\|_2
    =\min_{\mathcal S_i}\frac{\left\|\sum_{b\notin\mathcal S_i}m_{ib}(\bar{\mathbf{v}}_{ib}-\mathbf{o}_i)\right\|_2}{1-\sum_{b\notin\mathcal S_i}m_{ib}}.
    \label{eq:block_selection_objective}
\end{equation}
The numerator couples all discarded blocks inside a single norm, so the objective cannot be decomposed into independent per-block scores.
To measure each block's contribution to the output error, we apply the triangle inequality and obtain:
\begin{equation}
    \left\|\mathbf{o}_i-\widetilde{\mathbf{o}}_i\right\|_2
    \le\frac{\sum_{b\notin\mathcal S_i}m_{ib}\left\|\bar{\mathbf{v}}_{ib}-\mathbf{o}_i\right\|_2}{1-\sum_{b\notin\mathcal S_i}m_{ib}}.
    \label{eq:block_error_decomposition}
\end{equation}
Let \(p\) denote the overall mass threshold used to determine the retained block set.
Under a fixed threshold $p$, $1-\sum_{b\notin\mathcal S_i}m_{ib}$ remains almost constant around $p$.
Thus, minimizing the bound reduces to minimizing $\sum_{b\notin\mathcal S_i}m_{ib}\lVert\bar{\mathbf{v}}_{ib}-\mathbf{o}_i\rVert_2$.
We therefore adopt $i_{ib}=m_{ib}\lVert\bar{\mathbf{v}}_{ib}-\mathbf{o}_i\rVert_2$ as the importance proxy of block $b$, which combines attention mass and output deviation multiplicatively.

\textbf{An approximation to the block score.}
To obtain a proxy compatible with existing top-$p$ kernels, we apply a logarithmic transformation to $I_{ib}$ and express it as the sum of two terms.
To avoid the expensive computation of the exact block output, we approximate it with a first-order Taylor expansion, yielding the selection score $s_{ib}$ used for practical candidate ranking:
\begin{equation}
    s_{ib}
    =\frac{\bar{\mathbf{q}}_i\bar{\mathbf{k}}_b^\top}{\sqrt{d_h}}
    +\log\left\|\bar{\mathbf{v}}_{ib}-\mathbf{o}_i\right\|_2,
    \qquad
    \bar{\mathbf{v}}_{ib}\approx\bar{\mathbf{v}}_b+\frac{\bar{\mathbf{q}}_i}{\sqrt{d_h}}\mathbf{C}_b,
    \label{eq:ideal_log_proxy}
\end{equation}
where $\bar{\mathbf{v}}_b$ denotes the pooled value vector in block $b$,
$n_b=|\mathcal B_b|$, and
$\mathbf{C}_b=\frac{1}{n_b}\sum_{t\in\mathcal B_b}
(\mathbf{k}_t-\bar{\mathbf{k}}_b)^\top
(\mathbf{v}_t-\bar{\mathbf{v}}_b)$
is the key--value covariance of block $b$.
Appendix~\ref{app:derivations} provides the theoretical derivations, Appendix~\ref{app:algorithm_details} develops the selection score, and Appendix~\ref{app:experimental_setup} provides the experimental configuration.

\textbf{Two-stage selection.}
Computing the magnitude \(\bar{\mathbf{q}}_i\mathbf{C}_b/\sqrt{d_h}\) over all query--key block pairs requires \(\mathcal O(n^2)\) computation, which is unaffordable.
To avoid unnecessary computation on negligible blocks, we restrict this calculation to a small subset of potentially important blocks, which we call the candidate pool.
Therefore, we design a selector with two stages. Stage~1 selects the candidates, and Stage~2 further selects significant blocks using output deviation.
Given an overall mass target $p$, Stage~1 applies a conservative threshold $p_1\ge p$ to retain candidates while forming the coarse reference $\widehat{\mathbf{o}}_i=\sum_{b\le i}a_{ib}\bar{\mathbf{v}}_b$ as a prior for output $\mathbf{o}_i$, where $a_{ib}$ is the pooled attention weight of query $i$ over its visible blocks.
Stage~2 sets $p_2=p/p_1$ as its threshold, reranks candidates by the output-relative score, and applies $p_2$ to their renormalized attention weights, retaining nominal mass $p$ under the assumption that discarded blocks remain negligible.
Appendix~\ref{app:algorithm_details} provides the output reference construction and complete selection procedure.
Appendix~\ref{app:approximation_diagnostics} verifies the Stage~1 candidate-pool coverage, while Appendix~\ref{app:runtime_memory} analyzes the auxiliary memory overhead across sequence lengths.

\textbf{Output-relative score.}
To adjust the strength of the output-deviation term and better approximate the actual block importance $I_{ib}$, we introduce a scalar $\beta$ on the log term to balance attention mass and output deviation. This gives POEM's selection proxy for practical candidate reranking:
\begin{equation}
    s_{ib}=\frac{\bar{\mathbf{q}}_i\bar{\mathbf{k}}_b^\top}{\sqrt{d_h}}+\beta\log\left\|
    \bar{\mathbf{v}}_b+\frac{\bar{\mathbf{q}}_i}{\sqrt{d_h}}\mathbf{C}_b
    -\widehat{\mathbf{o}}_i
    \right\|_2.
    \label{eq:poem_selection_score}
\end{equation}
The first term estimates block-level attention strength, while the second
estimates the approximate block output's deviation from
$\widehat{\mathbf{o}}_i$. Following XAttention~\citep{xu2025xattention}, we
estimate the first term by antidiagonal scoring. These two magnitudes combine
to form our output-relative scorer.
\begin{wraptable}[9]{r}[-4pt]{0.38\textwidth}
  \centering
  \caption{Layer and logit output errors on LongBench, using Qwen3-8B.}
  \label{tab:metric_errors}
  \scriptsize
  \setlength{\tabcolsep}{3pt}
  \begin{tabular}{@{}lrrrr@{}}
    \toprule
    Method & \makecell[r]{Layer 5\\$(\times 10^{-4})$}
           & \makecell[r]{Layer 15\\$(\times 10^{-2})$}
           & Layer 35 & \makecell[r]{Final\\logits} \\
    \midrule
    Mass only & 3.27 & 1.24 & 5.88 & 2.86 \\
    STEM      & 3.27 & 1.23 & 5.88 & 2.85 \\
    \textbf{POEM}
      & \textbf{3.25} & \textbf{1.22}
      & \textbf{5.79} & \textbf{2.74} \\
    \bottomrule
  \end{tabular}
\end{wraptable}

Table~\ref{tab:metric_errors} compares the reconstruction errors of the three selection scores at the same density.
POEM consistently outperforms the mass-only baseline and STEM's value-aware proxy at all three evaluated layers and at the final logits, supporting the superiority of the output-relative proxy for block selection.
Appendix~\ref{app:experimental_setup} provides the detailed evaluation configuration, including the matched block density setting and definitions of the layer and logit reconstruction errors.
Figure~\ref{fig:ablation_studies}(d) examines the end-to-end impact of the output-relative proxy.
\par
\WFclear
\subsection{Hierarchical Error Decomposition}
\label{sec:hierarchical_error}
To keep the sparse output close to the dense output at the model level, we hierarchically decompose the prediction error and relate it to head-level reconstruction error.
Our overall objective is to select attention blocks that minimize the
discrepancy between dense and sparse prediction distributions under a fixed
block budget, which can be formalized as the following constrained objective:
\begin{equation}
    \min_{\mathcal S:\,|\mathcal S|=k}
    d_{\mathrm{KL}}\!\left(\mathbf{p}^d\,\|\,\mathbf{p}^s(\mathcal S)\right).
    \label{eq:predictive_goal}
\end{equation}
Here, $\mathcal{S}$ is the selected block set of size $k$, $\mathbf{p}^{d}$ and
$\mathbf{p}^{s}(\mathcal{S})$ are the prediction distributions under dense and sparse
attention, $d_{\mathrm{KL}}(\cdot\,\|\,\cdot)$ is the Kullback--Leibler (KL) divergence.
The objective is defined on the predictions, whereas block selection is made separately for each head, so we bridge this gap by decomposing the prediction error first into layer-level and then into head-level output errors.

\textbf{Step 1: Prediction to layer output.}
The objective in Eq.~\ref{eq:predictive_goal} is defined on the final predictions, but what block selection actually changes is the attention output of a single layer, so we begin by tracing how a perturbation of that output propagates to
the predictions.
Let $\mathbf{O}_\ell$ and $\widetilde{\mathbf{O}}_\ell$ denote the attention outputs of layer $\ell$ under dense and sparse attention, $\Delta\mathbf{O}_\ell := \widetilde{\mathbf{O}}_\ell - \mathbf{O}_\ell$ for the layer-output perturbation and $\boldsymbol{\epsilon}_\ell := \mathrm{vec}(\Delta\mathbf{O}_\ell)$ for its vectorization.
A first-order expansion of the downstream network gives the change in the final logits $\mathbf{z}$ as $\Delta\mathbf{z} = \mathbf{J}_{\ell\to z}\,\boldsymbol{\epsilon}_\ell +
\mathcal O\!\left(\|\boldsymbol{\epsilon}_\ell\|_2^2\right)$, where $\mathbf{J}_{\ell\to z}$ is the Jacobian from
the vectorized layer output to the logits. The perturbed logits shift the prediction
distribution from $\mathbf{p}^d = \mathrm{softmax}(\mathbf{z})$ to
$\mathbf{p}^s = \mathrm{softmax}(\mathbf{z} + \Delta\mathbf{z})$, and because the KL divergence attains its minimum at $\Delta\mathbf{z} = 0$, its expansion in $\Delta\mathbf{z}$ has no constant or first-order term,
\begin{equation}
d_{\mathrm{KL}}(\mathbf{p}^d \,\|\, \mathbf{p}^s)
= \tfrac{1}{2}\,\Delta\mathbf{z}^\top \mathbf{F}\,\Delta\mathbf{z} + \mathcal O\!\left(\|\Delta\mathbf{z}\|_2^3\right),
\qquad
\mathbf{F} := \mathrm{Diag}(\mathbf{p}^d) - \mathbf{p}^d (\mathbf{p}^d)^\top,
\label{eq:kl_logit_expansion}
\end{equation}
where $\mathbf{F}$ is the softmax Fisher matrix, which captures how strongly each direction of logit change perturbs the prediction distribution.
We further bound the quadratic form by the spectral norms of $\mathbf{F}$ and $\mathbf{J}_{\ell\to z}$, yielding the scalar upper bound
\begin{equation}
d_{\mathrm{KL}}(\mathbf{p}^d \,\|\, \mathbf{p}^s)
\le c_\ell\,\|\Delta\mathbf{O}_\ell\|_F^2 + \mathcal O\!\left(\|\Delta\mathbf{O}_\ell\|_F^3\right),
\qquad
c_\ell := \tfrac{1}{2}\,\|\mathbf{F}\|_2\,\|\mathbf{J}_{\ell\to z}\|_2^2,
\label{eq:prediction_reconstruction_bound}
\end{equation}
where the constant $c_\ell$ measures how sensitive the predictions are to this layer's output. The full derivation is provided in the theoretical appendix. The layer output still aggregates the contributions of all attention heads, so Step~2 splits this layer-level error
further into per-head output errors.

\textbf{Step 2: Layer to head.}
Block selection operates per attention head, so the layer-level bound must be
carried one step further, down to the heads that compose the layer output. Let
$\mathbf{o}_{\ell,h}$ and $\widetilde{\mathbf{o}}_{\ell,h}$ denote head $h$'s dense and sparse outputs
before projection, and let
$\Delta\mathbf{o}_{\ell,h} := \widetilde{\mathbf{o}}_{\ell,h} - \mathbf{o}_{\ell,h}$ be its output error.
Since Eq.~\ref{eq:sparse_attention_output} writes the layer output as a
sum of projected head outputs, the perturbation decomposes as
$\Delta\mathbf{O}_\ell = \sum_h \Delta\mathbf{o}_{\ell,h} \mathbf{W}_O^{(h)}$, and applying the triangle and Cauchy--Schwarz inequalities gives
\begin{equation}
\left\|\Delta\mathbf{O}_\ell\right\|_F^2
\le H \sum_h \left\|\mathbf{W}_O^{(h)}\right\|_2^2
\left\|\Delta\mathbf{o}_{\ell,h}\right\|_F^2,
\label{eq:layer_to_head_bound}
\end{equation}
where $H$ is the number of heads and $\mathbf{W}_O^{(h)}$ is the row block of the output
projection for head $h$.
This bound isolates each head's contribution through its projection norm and output error.
Combining this inequality with Eq.~\ref{eq:prediction_reconstruction_bound}
bounds the predictive divergence by a weighted sum of head-level errors.
The two factors motivate the two components of POEM.
Output-relative block selection (Section~\ref{sec:output_relative_block_selection})
targets the per-head error $\|\Delta\mathbf{o}_{\ell,h}\|_F$.
Headwise threshold calibration (Section~\ref{sec:headwise_budget}) accounts for
the unequal head sensitivities represented by the projection weights
$\|\mathbf{W}_O^{(h)}\|_2^2$ in the resulting predictive-divergence bound across all attention heads within the layer.

\subsection{Offline threshold calibration}
\label{sec:calibration}
\label{sec:headwise_budget}

\textbf{Error-ranked batch bisection.}
Offline calibration replaces the shared $p_2=p/p_1$ with a headwise threshold $p_{2,h}$, where $p_{2,h}$ is the Stage~2 threshold of head $h$.
Following GQA's KV grouping~\citep{ainslie2023gqa} and KV-cache compression methods~\citep{NEURIPS2025_57e0358a,rehg2024kv}, we jointly calibrate groups of heads that share KV projections.
To avoid calibrating groups individually, Algorithm~\ref{alg:calibration} in Appendix~\ref{app:threshold_calibration} sorts groups by their estimated output-distribution error and tests them in batches, starting from all groups.
Each trial lowers the batch thresholds.
If the score on the search set does not decrease, the reduction is accepted for the entire batch. Otherwise, the batch is bisected and the half with lower error is retried first in subsequent calibration.
The calibration ends when all groups have been processed or the predefined maximum number of steps is reached.

\noindent\begin{minipage}{\textwidth}
\subsection{POEM: Output-Relative Block-Sparse Attention}
\label{sec:two_stage_pipeline}
\begin{wrapfigure}[16]{r}{0.50\textwidth}
  \begin{minipage}{\linewidth}
    \setlength{\intextsep}{0pt}
    \begin{algorithm}[H]
  \captionsetup{font=scriptsize}
  \raggedright
  \caption{POEM: Output-Relative Block-Sparse Attention}
  \label{alg:poem}
  \scriptsize
  \algrenewcommand\alglinenumber[1]{\scriptsize #1:}
  \algrenewcommand\algorithmicindent{0.8em}
  \begin{algorithmic}[1]
    \State \textbf{Input:} $\mathbf{Q}$, $\mathbf{K}$, $\mathbf{V}$.
    \State \textbf{Parameters:} $p_1$, $\beta$, $\{p_{2,h}\}_h$, pooling stride $B$.
    \State \textbf{Output:} Sparse outputs $\widetilde{\mathbf{o}}_{i,h}$.
    \For{each head $h$ (using head-local $\mathbf{Q},\mathbf{K},\mathbf{V}$)}
      \State Cache block covariances $\mathbf{C}_b$.
      \For{each query block $i$}
        \State \textcolor{gray}{\# Stage 1}
        \State $\mathbf{a}_i\gets\operatorname{Pool}(\mathbf{Q},\mathbf{K},\mathrm{stride}=B)$. \Comment{\# Antidiagonal}
        \State $\mathcal C_i\gets\operatorname{TopP}(\mathbf{a}_i,p_1)$.
        \State $\widehat{\mathbf{o}}_i\gets\sum_{b\le i}a_{ib}\bar{\mathbf{v}}_b$.
        \State \textcolor{gray}{\# Stage 2}
        \State Compute $s_{ib}$ over $b\in\mathcal C_i$ using Eq.~\ref{eq:poem_selection_score}.
        \State ${Rerank}(s_{ib},\mathcal C_i)$.
        \State $\mathcal S_i\gets\operatorname{ThresholdP}(\mathcal C_i,p_{2,h})$. \Comment{\# Mass threshold}
        \State $\widetilde{\mathbf{o}}_{i,h}\gets\operatorname{Attention}(\mathbf{Q}_i,\mathbf{K},\mathbf{V};\mathcal S_i)$.
      \EndFor
    \EndFor
  \end{algorithmic}
    \end{algorithm}
  \end{minipage}
\end{wrapfigure}
We integrate the output-relative selection score and the offline threshold calibration together into POEM, a unified training-free framework.
Figure~\ref{fig:poem_framework} summarizes the execution order, and Algorithm~\ref{alg:poem} gives the detailed inference procedure.
\textbf{First}, POEM selects candidates by lightweight attention mass alone to quickly construct the candidate pool and compute the output reference $\widehat{\mathbf{o}}_i$ for subsequent output-relative ranking.
\textbf{Second}, POEM computes the output-relative score in Eq.~(\ref{eq:poem_selection_score}) within the candidate pool, which combines attention mass with output deviation to account for each block's importance in reconstructing the output.
For a head $h$, POEM reranks the blocks in the candidate pool by the output-relative score and applies the calibrated threshold $p_{2,h}$ to the reordered sequence, where $p_{2,h}$ is the calibrated threshold.
\par
\WFclear
\end{minipage}
\par
\medskip

\section{Experiments}
\label{sec:experiment}
\textbf{Baselines.}
We compare POEM with dense FlashAttention-2~\citep{dao2024flashattention} and four training-free sparse prefilling baselines under the same FlashAttention-2 backend: \textbf{MInference}~\citep{jiang2024minference} (official sparse-head configuration), \textbf{FlexPrefill}~\citep{lai2025flexprefill} ($B=128$, $\gamma=0.9$, $\tau=0.1$), \textbf{XAttention}~\citep{xu2025xattention} ($\tau=0.9$), and \textbf{STEM}~\citep{niu2026stem} ($B=128$, $\mu=0.7$, $\beta=0.2$, $k_{\mathrm{start}}=0.2N_{\mathrm{blk}}$ for 8--16K inputs and $0.1N_{\mathrm{blk}}$ beyond 16K, with a minimum budget of 54 blocks).
We use each baseline's released or recommended configuration without task-specific retuning.
SpargeAttention~\citep{zhang2025spargeattention} is excluded because its online \(PV\) skipping falls outside this shared-backend block-selection comparison and its controlled evaluation scope.

\textbf{Implementation details.}
We implement all experiments in PyTorch on an NVIDIA A100 GPU.
After block selection, FlashAttention-2 computes attention over the retained blocks.
Following XAttention and STEM, we retain local and sink blocks for stability: four local and four sink blocks for Llama-3.1-8B~\citep{grattafiori2024llama}, and one of each for Qwen3-8B/14B~\citep{yang2025qwen3}.
In Stage~1, we estimate block attention mass from low-precision query and key vectors to select the candidate pool.
In Stage~2, we compute output-relative scores within the candidate pool.
We set the output-relative deviation coefficient to $\beta=0.3$ for Qwen3-14B and Llama-3.1-8B, and to $\beta=0.2$ for Qwen3-8B.
For Qwen-family models, we use YaRN~\citep{peng2026yarnefficientcontextwindow} to extend their context windows to 128K, enabling long-context evaluation.
We set the overall mass target to $p=0.9$ and use the Stage~1 threshold $p_1=0.94$, yielding $p_2=p/p_1$ for Stage~2.

\begin{table}[h]
  \centering
  \caption{LongBench and RULER results on three models. Best in bold, second best underlined.}
  \label{tab:main_results}
  \scriptsize
  \setlength{\tabcolsep}{2.0pt}
  \renewcommand{\arraystretch}{1.08}
  \resizebox{\textwidth}{!}{%
  \begin{tabular}{ll*{14}{c}}
    \toprule
    \multirow{2}{*}{Model} & \multirow{2}{*}{Method}
      & \multicolumn{7}{c}{LongBench}
      & \multicolumn{7}{c}{RULER} \\
    \cmidrule(lr){3-9}\cmidrule(lr){10-16}
      & & S-Doc & M-Doc & Summ. & FSL & Synth. & Code & \textbf{Avg.}
      & 4K & 8K & 16K & 32K & 64K & 128K & \textbf{Avg.} \\
    \midrule
    \multirow{6}{*}{\makecell[l]{Llama-3.1-8B}}
      & Dense & 28.57 & 25.75 & 25.45 & 63.61 & 65.83 & 33.06 & 40.38
              & 95.26 & 94.29 & 93.58 & 87.99 & 84.85 & 77.87 & 88.97 \\
      & MInference & 28.38 & 25.44 & 25.28 & 63.31 & 65.12 & 31.16 & \underline{39.78}
                   & 95.43 & 94.25 & 93.44 & 87.56 & 83.21 & 71.51 & 87.57 \\
      & FlexPrefill & 24.25 & 22.61 & 25.14 & 62.24 & 19.56 & 35.01 & 31.47
                    & 93.67 & 93.63 & 93.84 & 87.90 & 84.83 & 72.43 & \underline{87.71} \\
      & XAttention & 27.51 & 23.55 & 25.51 & 63.82 & 55.83 & 35.77 & 38.66
                   & 95.09 & 94.39 & 93.27 & 88.95 & 82.45 & 70.09 & 87.38 \\
      & STEM & 27.20 & 24.33 & 25.41 & 62.56 & 64.32 & 32.74 & 39.43
             & 95.27 & 94.50 & 93.12 & 88.21 & 82.94 & 70.33 & 87.40 \\
      \rowcolor{black!10}
      & POEM & 26.82 & 23.98 & 25.44 & 64.04 & 65.71 & 35.00 & \textbf{40.16} & 95.01 & 94.32 & 93.20 & 88.79 & 82.74 & 72.38 & \textbf{87.74} \\
    \midrule
    \multirow{6}{*}{\makecell[l]{Qwen3-8B}}
      & Dense & 14.48 & 13.12 & 22.14 & 60.59 & 66.51 & 12.54 & 31.56
              & 95.18 & 92.51 & 89.81 & 86.78 & 79.55 & 75.29 & 86.52 \\
      & MInference & 13.97 & 12.73 & 21.45 & 58.06 & 65.75 & 12.02 & 30.66
                   & 94.83 & 92.55 & 89.41 & 85.68 & 76.12 & 70.89 & \underline{84.91} \\
      & FlexPrefill & 12.87 & 12.11 & 21.47 & 61.44 & 42.87 & 14.64 & 27.57
                    & 90.94 & 87.85 & 87.45 & 85.12 & 78.94 & 73.98 & 84.05 \\
      & XAttention & 13.36 & 12.77 & 21.38 & 58.72 & 63.54 & 14.42 & 30.70
                   & 94.56 & 90.80 & 87.92 & 84.54 & 78.02 & 71.70 & 84.59 \\
      & STEM & 14.13 & 13.00 & 21.39 & 57.79 & 65.62 & 12.46 & \underline{30.73}
             & 94.67 & 90.43 & 87.61 & 85.09 & 77.42 & 71.24 & 84.41 \\
      \rowcolor{black!10}
      & POEM & 13.65 & 13.12 & 21.62 & 58.56 & 62.97 & 15.08 & \textbf{30.83} & 94.27 & 91.32 & 88.11 & 85.30 & 77.92 & 72.71 & \textbf{84.94} \\
    \midrule
    \multirow{6}{*}{\makecell[l]{Qwen3-14B}}
      & Dense & 15.55 & 12.62 & 21.92 & 62.41 & 66.83 & 53.30 & 38.77
              & 95.61 & 93.83 & 93.27 & 93.16 & 90.45 & 84.01 & 91.72 \\
      & MInference & 15.50 & 12.51 & 21.76 & 62.04 & 65.58 & 51.40 & 38.13
                   & 95.34 & 93.28 & 91.16 & 90.11 & 87.44 & 80.46 & \underline{89.63} \\
      & FlexPrefill & 15.09 & 13.05 & 21.77 & 61.15 & 48.12 & 50.95 & 35.02
                    & 93.43 & 92.74 & 92.30 & 91.13 & 87.41 & 80.64 & 89.60 \\
      & XAttention & 15.37 & 13.11 & 21.82 & 60.94 & 65.42 & 52.67 & 38.22
                   & 95.19 & 93.02 & 91.23 & 89.89 & 87.36 & 80.33 & 89.50 \\
      & STEM & 15.35 & 12.82 & 21.71 & 60.86 & 66.92 & 51.77 & \underline{38.24}
             & 95.78 & 92.85 & 91.08 & 89.92 & 87.22 & 80.01 & 89.48 \\
      \rowcolor{black!10}
      & POEM & 15.36 & 13.21 & 21.83 & 62.12 & 65.92 & 53.61 & \textbf{38.68} & 95.23 & 93.12 & 91.60 & 89.95 & 87.64 & 80.51 & \textbf{89.68} \\
    \bottomrule
  \end{tabular}
  }
\end{table}

\textbf{Comparison on LongBench and RULER.}
Table~\ref{tab:main_results} compares POEM with dense attention and four training-free sparse baselines, including MInference,
FlexPrefill, XAttention,
and STEM, on LongBench
and RULER, covering models from
different families and at different scales: Llama-3.1-8B
and Qwen3-8B/14B.
On LongBench, POEM ranks first among all sparse methods on every model,
narrowing the gap to dense attention to as little as 0.09 points on Qwen3-14B.
On RULER, POEM achieves the best average score across all three evaluated models.
Across the two benchmarks, POEM is the only sparse method that consistently matches or outperforms every baseline on all model configurations.
We attribute this robustness to the output-relative block scoring and head-wise threshold calibration, which together better reconstruct the dense attention output.

\begin{figure}[H]
  \centering
  \includegraphics[width=\textwidth]{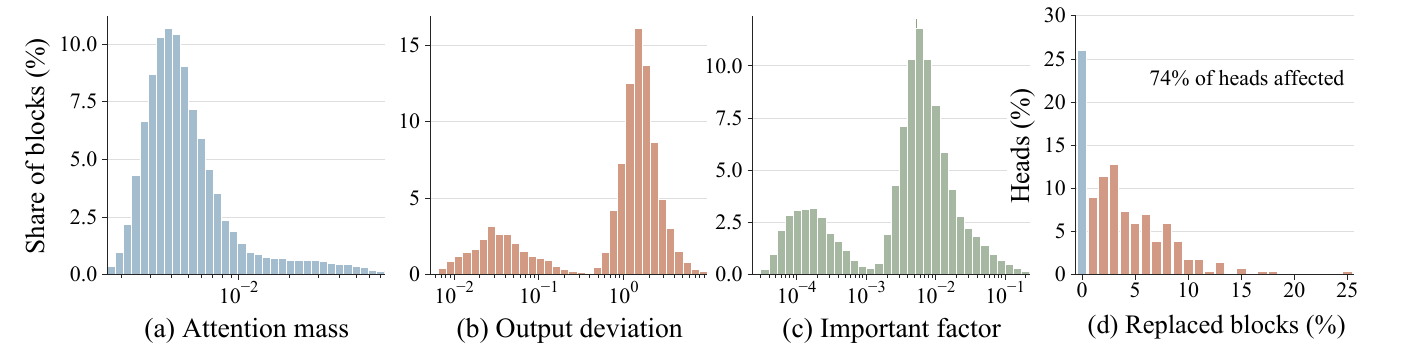}
  \vspace{-1.2em}
  \caption{
   (a)~Distribution of exact block attention mass $m_{ib}$.
  (b)~Distribution of output deviation $\lVert\bar{v}_{ib}-o_i\rVert_2$.
  (c)~Distribution of exact importance $I_{ib}=m_{ib}\lVert\bar{v}_{ib}-o_i\rVert_2$.
  (d)~Percentage of blocks replaced per head after output-relative reranking, 74\% of heads are affected.}
  \label{fig:metric_value_reranking}
\end{figure}

\textbf{Effect of output-relative reranking.}
To examine whether output deviation provides a selection signal beyond attention mass, we conduct a block-level diagnostic using exact quantities computed with the dense output $o_i$.
Figure~\ref{fig:metric_value_reranking} (a) shows the distribution of exact block attention mass $m_{ib}$, which is concentrated near zero and already provides strong discrimination between blocks.
Figure~\ref{fig:metric_value_reranking} (b) plots the output deviation $\lVert\bar v_{ib}-o_i\rVert_2$, which is divided into two distinct modes. The low-deviation mode corresponds to blocks whose values already lie close to the dense output, whereas the high-deviation mode contains blocks that would shift the output substantially if removed. Figure~\ref{fig:metric_value_reranking} (c) displays the importance factor $I_{ib}=m_{ib}\lVert\bar v_{ib}-o_i\rVert_2$, which POEM's practical score in Eq.~\ref{eq:poem_selection_score} approximates.
Incorporating the deviation reshapes the distribution relative to mass alone: blocks that rank high by mass but contribute values close to the output are lowered, while  blocks with moderate mass but large deviation gain relative importance.
As Figure~\ref{fig:metric_value_reranking} (d) shows, this change of distribution alters the selected block set in more than 74\% of all heads, indicating that output deviation provides an important complementary signal for reliable block selection.

\begin{wraptable}{r}{0.35\textwidth}
  \centering
  \captionsetup{position=top}
  \caption{PG19 ablation at 32k with 0.1 density on Qwen3-8B.}
  \label{tab:cov_deviation_ablation}
  \begin{tabular*}{\linewidth}{@{\extracolsep{\fill}}lcc@{}}
    \toprule
    Variant & Word PPL & \makecell{$\Delta$ PPL} \\
    \midrule
    POEM & 43.65 & $+0.00$ \\
    w/o cov. & 44.01 & $+0.36$ \\
    w/o output & 43.95 & $+0.30$ \\
    \bottomrule
  \end{tabular*}
\end{wraptable}
\textbf{Module ablation.}
Table~\ref{tab:cov_deviation_ablation} separately ablates covariance correction and the output reference on Qwen3-8B.
Removing either component worsens perplexity, confirming that covariance improves block-output estimation, while the output reference enables output-relative deviation.
Table~\ref{tab:module_ablation_128k} presents the module ablation on
RULER at 128K context for Llama-3.1-8B, where we gradually
add each component of POEM to a baseline that selects blocks by mass alone.
Incorporating the output deviation into the block score yields a broad improvement, lifting the average from 69.94 to 72.14 at almost the same budget (19.50\% vs.\ 19.56\%).
Further adding threshold calibration gives this score a slight increase (72.38) while reducing the density from 19.56\% to 18.53\%, confirming that calibration can leverage differences in head sensitivity.

\begin{table}[!ht]
  \centering
  \captionsetup{position=top}
  \caption{POEM module ablation on RULER at 128K. Best results are in bold, and second-best results are underlined.}
  \label{tab:module_ablation_128k}
  \scriptsize
  \setlength{\tabcolsep}{2.5pt}
  \renewcommand{\arraystretch}{1.1}
  \resizebox{\textwidth}{!}{%
  \begin{tabular}{l*{15}{c}}
    \toprule
    \multirow{2}{*}{Method}
      & \multicolumn{8}{c}{Retrieval}
      & MHT
      & \multicolumn{2}{c}{Agg.}
      & \multicolumn{2}{c}{QA}
      & Overall & \multirow{2}{*}{Budget} \\
    \cmidrule(lr){2-9}\cmidrule(lr){10-10}\cmidrule(lr){11-12}\cmidrule(lr){13-14}\cmidrule(lr){15-15}
      & S1 & S2 & S3 & MK1 & MK2 & MK3 & MV & MQ
      & VT & CWE & FWE & QA1 & QA2 & Avg. & \\
    \midrule
    Baseline & \textbf{100.00} & \textbf{99.00} & \textbf{100.00} & 92.00 & 62.00 & 43.00 & \underline{93.75} & 97.75 & 68.20 & \underline{0.10} & \underline{73.00} & 44.42 & \underline{36.00} & 69.94 & \underline{19.50}\% \\
    + output deviation & \textbf{100.00} & \textbf{99.00} & \textbf{100.00} & \underline{93.00} & \underline{69.00} & \textbf{48.00} & \textbf{95.75} & \underline{98.25} & \underline{71.00} & \underline{0.10} & \textbf{73.67} & \underline{52.08} & \textbf{38.00} & \underline{72.14} & 19.56\% \\
    \textbf{+ threshold calibration} & \textbf{100.00} & \underline{98.00} & \textbf{100.00} & \textbf{94.00} & \textbf{70.00} & \underline{46.00} & \textbf{95.75} & \textbf{98.50} & \textbf{74.80} & \textbf{0.20} & \underline{73.00} & \textbf{54.75} & \underline{36.00} & \textbf{72.38} & \textbf{18.53}\% \\
    \bottomrule
  \end{tabular}%
  }
\end{table}

\begin{figure}[!ht]
  \centering
  \includegraphics[width=\textwidth]{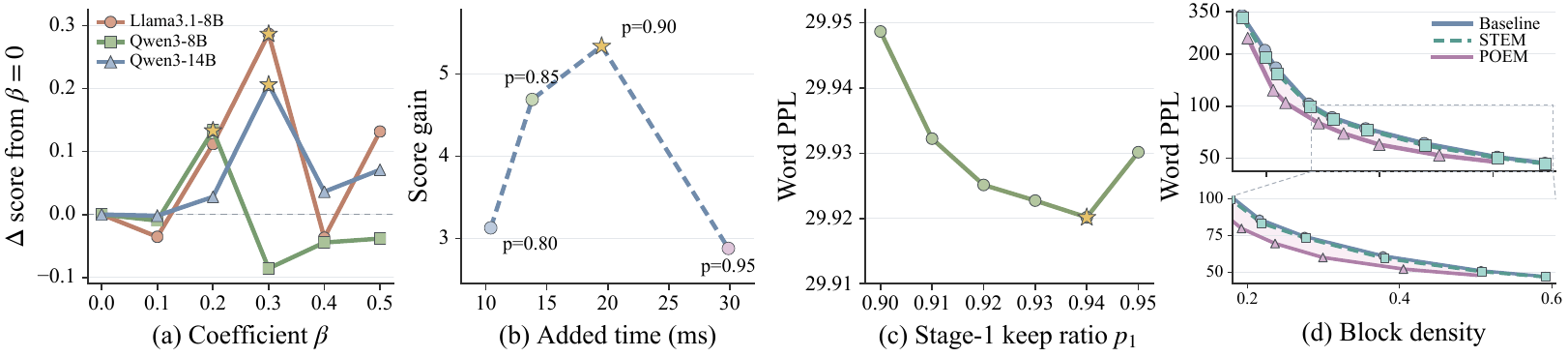}
  \caption{POEM ablations: (a)~output-relative weight. (b)~ score--time trade-off.
  (c)~Stage-1 mass threshold. (d)~PG19 perplexity vs density for different scoring proxies.}
  \label{fig:ablation_studies}
\end{figure}

\color{black}
\textbf{Ablation studies.}
We conduct ablation studies on the output-relative coefficient $\beta$, total threshold $p$, stage~1 threshold $p_1$, and different scoring proxies on LongBench, RULER, and PG19~\citep{raecompressive2019}. Figures~\ref{fig:ablation_studies}(b)--(d) use Llama 3.1-8B.
Figure~\ref{fig:ablation_studies} (a) shows that performance peaks at $\beta=0.3$ for Qwen 3-14B and Llama 3-8B, and at $0.2$ for Qwen 3-8B.
Larger values degrade performance, suggesting that moderate deviation weighting improves selection but excessive weighting introduces noise from the attention mass.
Figure~\ref{fig:ablation_studies} (b) plots the gain against added latency as $p$ increases on RULER.
The gain increases from $p=0.8$ to $p=0.9$, but drops at $p=0.95$ despite the higher latency, showing that $p=0.9$ provides the best trade-off among the evaluated settings. Figure~\ref{fig:ablation_studies} (c) varies the Stage-1 threshold $p_1$ under $p = 0.9$.
PG19 perplexity decreases as $p_1$ grows from 0.90 to 0.94, since a larger candidate pool gives the output-relative proxy more room to improve upon mass-only selection. Beyond 0.94, however, the coarse attention score used in Stage-1 admit increasingly unreliable blocks into the pool, and because the final retention threshold $p$ is held fixed, these blocks displace trustworthy candidates and thus worsen the perplexity.
Figure~\ref{fig:ablation_studies} (d) confirms our output-relative proxy's consistent superiority over mass-only and STEM scoring at matched PG19 densities. Details for all ablations are provided in Appendix~\ref{app:ablation_setup}.

\begin{figure}[!ht]
  \centering
  \includegraphics[width=0.9\textwidth]{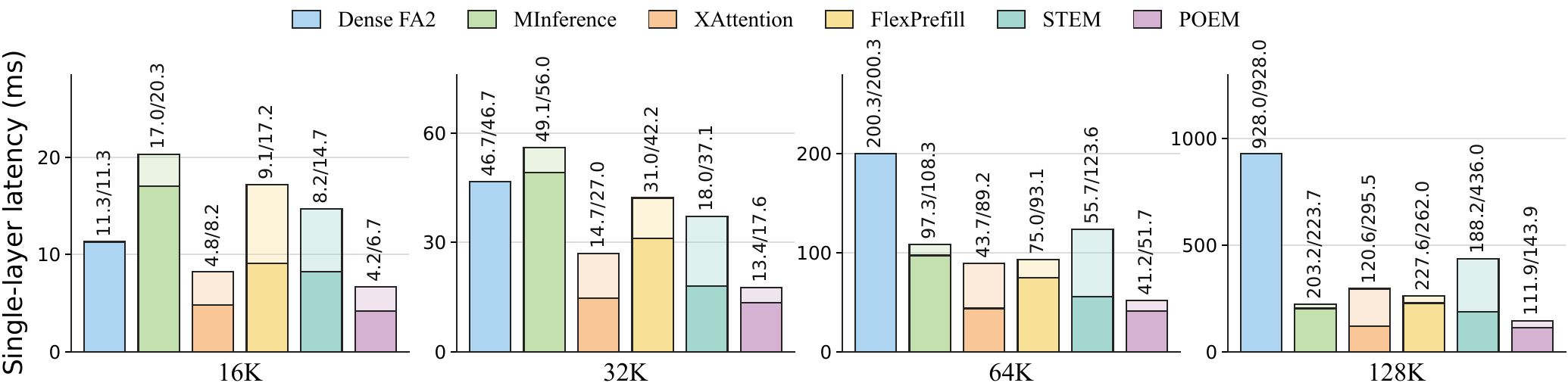}
  \vspace{-1em}
  \caption{Attention latency (kernel/total). Dark/light segments show kernel/selector time. POEM is fastest throughout, reaching $6.5\times$ speedup over dense FA2 at 128K.}
  \label{fig:efficiency_breakdown}
\end{figure}

\textbf{Total efficiency analysis.}
Figure~\ref{fig:efficiency_breakdown} compares the total attention run time of five state-of-the-art methods against the FlashAttention-2 baseline, from 16k to 128k, on Qwen3-8B.
In these methods, POEM maintains fastest in all the length.
At the length of 128k tokens, POEM even accelerates Flashattention-2 by 6.5$\times$, from 928.0 ms to 143.9 ms, saving a massive amount of time.

\begin{figure}[!ht]
  \centering
  \includegraphics[width=\textwidth]{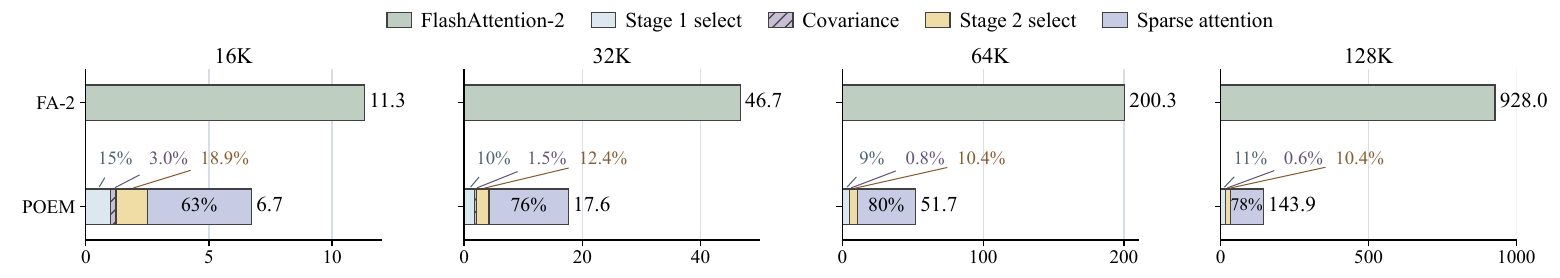}
  \vspace{-1em}
  \caption{Latency breakdown of POEM. Selection overhead remains modest across context lengths, while sparse attention dominates the execution time at long contexts.}
  \label{fig:poem_stage_timing}
\end{figure}

\textbf{Latency breakdown of POEM.}
Furthermore, we analysis the detailed attention execution composition of POEM in Figure~\ref{fig:poem_stage_timing}.
The results show that the two-stage selection process of POEM is modest.
As length increasing, the selection process consistently accounts for only a small part of the total time of attention execution.
The attention kernel time takes the majority, accounting for around 80\% of the total times when lengths exceeds 64k. We conduct this study on Qwen3-8B.

\section{Conclusion}
\label{sec:conclusion}

We showed that the output error of sparse attention grows not only with the attention mass of the removed blocks but also with their output deviation from the dense attention output.
POEM therefore scores blocks by both quantities and achieves the best accuracy among sparse attention methods with up to a $6.5\times$ attention speedup over FlashAttention-2 at 128K tokens.
Future work could estimate block outputs more accurately before attention and extend the output-relative criterion to KV cache eviction during decoding and to trainable sparse attention.
\subsection*{AI Use Statement}

Generative AI tools were used to assist with drafting and language editing of
this manuscript, including improvements to clarity, grammar, and organization.
They were not treated as sources of scientific evidence: technical statements,
citations, equations, and reported results were checked by the authors against
the cited literature, mathematical derivations, implementation, and
experimental records.  All AI-assisted text was reviewed and revised by the
authors.  The authors take full responsibility for the final content of the
paper, including all text, claims, and artifacts produced with the assistance
of generative AI.

\subsection*{Ethics Statement}

This work studies the efficiency of long-context language-model inference
using publicly available pretrained models and established benchmark datasets;
it does not involve human participants, newly collected personal data, or
interaction with deployed users.  Because POEM approximates dense attention,
it can change model outputs in ways that are not fully captured by aggregate
benchmark scores, potentially increasing factual errors or amplifying biases
already present in the underlying models and data.  We therefore do not
recommend deploying the method in safety-critical or high-stakes settings
without application-specific evaluation and safeguards.  Although sparse
attention may reduce the memory and computation required for inference, the
experiments themselves consume GPU resources.  Users of the method remain
responsible for complying with the licenses and usage conditions of the
underlying models and datasets.

\subsection*{Reproducibility Statement}

Section~\ref{sec:method} specifies the two-stage POEM algorithm, its
output-relative objective, selection metric, and hyperparameters.  The models,
datasets, baselines, hardware, attention implementation, evaluation framework,
and shared evaluation settings are documented in
Section~\ref{sec:experiment} and Appendix~\ref{app:models_datasets}.  Appendix~\ref{app:derivations}
provides the detailed derivation of the sparse-attention output error and the
pooled approximation used by POEM.  The experiments use publicly available
model checkpoints and benchmarks, and all methods are evaluated using the same
examples and decoding settings.  To support independent reproduction, the
implementation, evaluation commands, configuration files, software versions,
and random seeds will be included in the anonymized supplementary material.


\bibliography{iclr2027_conference}
\bibliographystyle{iclr2027_conference}

\appendix
\numberwithin{equation}{section}

\section{Theoretical Analysis and Proofs}
\label{app:derivations}
This section separates the theoretical steps in POEM's motivation: a local
upper bound relating predictive divergence to layer-output error, an exact
identity for sparse-attention
output error, and the first-order approximation that makes block scoring
tractable. Algorithmic conventions, experimental protocols, and efficiency
accounting are deferred to later sections so that each derivation states only
the claim, assumptions, proof, and scope that it needs.
Throughout the appendices, bold uppercase letters denote matrices, bold
lowercase letters denote vectors, and plain lowercase letters denote
scalars. Calligraphic letters denote sets or structured collections, and
operator names retain their conventional notation. An index identifying
a block, head, or token does not remove boldface when the indexed object
is still a matrix or vector; scalar coordinates are written in plain type.
\subsection{Derivation of the Prediction-to-Layer-Output Bound}
\label{app:prediction_bound}
The objective in Eq.~\ref{eq:predictive_goal} is defined on the final
prediction distribution, whereas sparse block selection directly changes an
intermediate attention output. This derivation connects these two quantities.
We first propagate a layer-output perturbation to the final logits and then
expand the predictive KL divergence around the dense logits. The resulting
upper bound relates final prediction error to layer-output reconstruction
error, providing the first step of the hierarchical decomposition in
Section~\ref{sec:hierarchical_error}.

\paragraph{Proposition (prediction-to-layer-output bound).}
For a fixed query block $i$, suppose only layer $\ell$ is perturbed and the
downstream map from its projected attention output to the logits is locally
twice differentiable. Then, in a neighborhood of the dense computation path,
the predictive KL divergence satisfies Eq.~\ref{eq:app_prediction_bound};
in particular, its leading term is at most quadratic in the Frobenius norm of
the layer-output perturbation.

\paragraph{Assumptions.}
We apply sparse attention only at layer $\ell$ and keep all other layers dense.
The downstream logit map is twice continuously differentiable, with bounded
first and second derivatives near the dense output, and the perturbation is
small enough for the Taylor expansions below. For query block $i$,
define the attention-output perturbation and its vectorization as
\[
    \Delta \mathbf{O}_{\ell,i}
    =\widetilde{\mathbf{O}}_{\ell,i}-\mathbf{O}_{\ell,i},
    \qquad
    \boldsymbol{\varepsilon}_{\ell,i}=\operatorname{vec}(\Delta \mathbf{O}_{\ell,i}),
\]
so that
\[
    \lVert\boldsymbol{\varepsilon}_{\ell,i}\rVert_2
    =\lVert\Delta \mathbf{O}_{\ell,i}\rVert_F.
\]
For compactness, write $\mathbf{x}_{\ell,i}=\operatorname{vec}(\mathbf{O}_{\ell,i})$ and let
$f_{\ell\to z}$ denote the downstream network from this vectorized attention
output to the final logits.  Thus,
\[
    \mathbf{z}_i^d=f_{\ell\to z}(\mathbf{x}_{\ell,i}),
    \qquad
    \mathbf{z}_i^s=f_{\ell\to z}(\mathbf{x}_{\ell,i}+\boldsymbol{\varepsilon}_{\ell,i}),
    \qquad
    \Delta \mathbf{z}_i=\mathbf{z}_i^s-\mathbf{z}_i^d.
\]
\paragraph{Proof.}
\emph{From output perturbations to logits.}
Writing $z_{ic}$ for coordinate $c$ of $\mathbf{z}_i$, Taylor's theorem gives
\begin{equation}
    \Delta z_{ic}
    =\nabla f_{\ell\to z,c}(\mathbf{x}_{\ell,i})^\top
      \boldsymbol{\varepsilon}_{\ell,i}
     +\frac{1}{2}\boldsymbol{\varepsilon}_{\ell,i}^\top
      \mathbf{H}_{\ell\to z,c}(\mathbf{x}_{\ell,i}+\theta_c\boldsymbol{\varepsilon}_{\ell,i})
      \boldsymbol{\varepsilon}_{\ell,i},
    \qquad \theta_c\in(0,1),
    \label{eq:app_logit_coordinate_taylor}
\end{equation}
where $\mathbf{H}_{\ell\to z,c}$ is the Hessian of the $c$th logit with respect to
the vectorized attention output.  If these Hessians are bounded in a
neighborhood of the dense path, stacking
Eq.~\ref{eq:app_logit_coordinate_taylor} over all logits gives
\begin{equation}
    \Delta \mathbf{z}_i
    =\mathbf{J}_{\ell\to z,i}\boldsymbol{\varepsilon}_{\ell,i}
     +\mathbf{r}_{\ell,i},
    \qquad
    \lVert \mathbf{r}_{\ell,i}\rVert_2
    =\mathcal O\!\left(\lVert\boldsymbol{\varepsilon}_{\ell,i}\rVert_2^2\right),
    \label{eq:app_logit_first_order}
\end{equation}
where $\mathbf{J}_{\ell\to z,i}$ is the Jacobian of $f_{\ell\to z}$ evaluated on the
dense computation path.  For notational convenience, the Jacobian is taken
with respect to the vectorized attention output.

\emph{The local KL expansion.}
Let $\mathbf{z}_i^d$ be the dense logits and define
\[
    \mathbf{p}_i^d=\operatorname{softmax}(\mathbf{z}_i^d),
    \qquad
    \mathbf{p}_i^s=\operatorname{softmax}(\mathbf{z}_i^d+\Delta \mathbf{z}_i),
    \qquad
    \Delta l_i^{\mathrm{pred}}
    =d_{\mathrm{KL}}(\mathbf{p}_i^d\,\|\,\mathbf{p}_i^s).
\]
To calculate this KL divergence explicitly, define the log-partition function
\begin{equation}
    a(\mathbf{z})=\log\!\left(\sum_c \exp(z_c)\right).
\end{equation}
Because $\log[\operatorname{softmax}(\mathbf{z})]_c=z_c-a(\mathbf{z})$, we have
\begin{align}
    \Delta l_i^{\mathrm{pred}}
    &=\sum_c p_{ic}^d
      \left[\log p_{ic}^d-\log p_{ic}^s\right] \nonumber\\
    &=\sum_c p_{ic}^d
      \left[-\Delta z_{ic}
      +a(\mathbf{z}_i^d+\Delta \mathbf{z}_i)-a(\mathbf{z}_i^d)\right] \nonumber\\
    &=a(\mathbf{z}_i^d+\Delta \mathbf{z}_i)-a(\mathbf{z}_i^d)
      -(\mathbf{p}_i^d)^\top\Delta \mathbf{z}_i,
    \label{eq:app_kl_log_partition}
\end{align}
where the last equality uses $\sum_c p_{ic}^d=1$. Differentiating the same
log-partition expression once and then once more keeps the calculation in one
chain:
\begin{equation}
\begin{alignedat}{2}
    \frac{\partial a(\mathbf{z})}{\partial z_c}
    &=\frac{\exp(z_c)}{\sum_j\exp(z_j)}=p_c,
    &\qquad
    \frac{\partial^2 a(\mathbf{z})}{\partial z_c\partial z_d}
    &=p_c\bigl(\mathbb I[c=d]-p_d\bigr),\\
    \nabla a(\mathbf{z}_i^d)
    &=\mathbf{p}_i^d,
    &
    \nabla^2 a(\mathbf{z}_i^d)
    &=\operatorname{Diag}(\mathbf{p}_i^d)-\mathbf{p}_i^d(\mathbf{p}_i^d)^\top
      =\mathbf{F}_i.
\end{alignedat}
\label{eq:app_log_partition_derivatives}
\end{equation}
Expanding the log-partition function in
Eq.~\ref{eq:app_kl_log_partition} therefore gives
\begin{align}
    a(\mathbf{z}_i^d+\Delta \mathbf{z}_i)
    &=a(\mathbf{z}_i^d)+(\mathbf{p}_i^d)^\top\Delta \mathbf{z}_i
      +\frac{1}{2}\Delta \mathbf{z}_i^\top \mathbf{F}_i\Delta \mathbf{z}_i
      +\mathcal O\!\left(\lVert\Delta \mathbf{z}_i\rVert_2^3\right).
    \label{eq:app_log_partition_taylor}
\end{align}
The constant terms cancel in Eq.~\ref{eq:app_kl_log_partition}, and its
linear term cancels the explicit term $-(\mathbf{p}_i^d)^\top\Delta \mathbf{z}_i$.  Hence,
\begin{equation}
    \Delta l_i^{\mathrm{pred}}
    =\frac{1}{2}\Delta \mathbf{z}_i^\top \mathbf{F}_i\Delta \mathbf{z}_i
    +\mathcal O\!\left(\lVert\Delta \mathbf{z}_i\rVert_2^3\right),
    \label{eq:app_kl_second_order}
\end{equation}
This calculation also shows directly why the first-order term vanishes and
why the quadratic term is governed by the softmax Fisher matrix $\mathbf{F}_i$.

\emph{Bounding the quadratic term.}
Substituting Eq.~\ref{eq:app_logit_first_order} into
Eq.~\ref{eq:app_kl_second_order}, the quadratic term can be expanded as
\begin{align}
    \frac{1}{2}\Delta \mathbf{z}_i^\top \mathbf{F}_i\Delta \mathbf{z}_i
    &=\frac{1}{2}
      (\mathbf{J}_{\ell\to z,i}\boldsymbol{\varepsilon}_{\ell,i}+\mathbf{r}_{\ell,i})^\top
      \mathbf{F}_i
      (\mathbf{J}_{\ell\to z,i}\boldsymbol{\varepsilon}_{\ell,i}+\mathbf{r}_{\ell,i}) \nonumber\\
    &=\frac{1}{2}\boldsymbol{\varepsilon}_{\ell,i}^\top
      \mathbf{J}_{\ell\to z,i}^\top \mathbf{F}_i\mathbf{J}_{\ell\to z,i}
      \boldsymbol{\varepsilon}_{\ell,i}
      +\boldsymbol{\varepsilon}_{\ell,i}^\top \mathbf{J}_{\ell\to z,i}^\top
       \mathbf{F}_i\mathbf{r}_{\ell,i}
      +\frac{1}{2}\mathbf{r}_{\ell,i}^\top \mathbf{F}_i\mathbf{r}_{\ell,i}.
    \label{eq:app_quadratic_substitution}
\end{align}
Locally, $\mathbf{J}_{\ell\to z,i}$ and $\mathbf{F}_i$ are bounded, while
$\lVert \mathbf{r}_{\ell,i}\rVert_2=\mathcal O(\lVert\boldsymbol{\varepsilon}_{\ell,i}\rVert_2^2)$.
The second and third terms in
Eq.~\ref{eq:app_quadratic_substitution} are therefore respectively
$\mathcal O(\lVert\boldsymbol{\varepsilon}_{\ell,i}\rVert_2^3)$ and
$\mathcal O(\lVert\boldsymbol{\varepsilon}_{\ell,i}\rVert_2^4)$.  Moreover,
$\lVert\Delta \mathbf{z}_i\rVert_2=\mathcal O(\lVert\boldsymbol{\varepsilon}_{\ell,i}\rVert_2)$,
so the remainder in Eq.~\ref{eq:app_kl_second_order} is also
$\mathcal O(\lVert\boldsymbol{\varepsilon}_{\ell,i}\rVert_2^3)$.  Collecting these terms
yields
\begin{equation}
    \Delta l_i^{\mathrm{pred}}
    =\frac{1}{2}\boldsymbol{\varepsilon}_{\ell,i}^\top
      \mathbf{J}_{\ell\to z,i}^\top \mathbf{F}_i\mathbf{J}_{\ell\to z,i}\boldsymbol{\varepsilon}_{\ell,i}
    +\mathcal O\!\left(\lVert\boldsymbol{\varepsilon}_{\ell,i}\rVert_2^3\right).
    \label{eq:app_output_to_prediction}
\end{equation}
This expression connects the layer-output perturbation to the final
predictive divergence through the downstream Jacobian and the softmax Fisher
matrix.
Assume locally that
\[
    \lVert \mathbf{J}_{\ell\to z,i}\rVert_2\le l_{\ell,i},
    \qquad
    \lVert \mathbf{F}_i\rVert_2\le\kappa_i.
\]
Since $\mathbf{F}_i$ is positive semidefinite, the leading quadratic term satisfies
\begin{align}
    \boldsymbol{\varepsilon}_{\ell,i}^\top
      \mathbf{J}_{\ell\to z,i}^\top \mathbf{F}_i\mathbf{J}_{\ell\to z,i}
      \boldsymbol{\varepsilon}_{\ell,i}
    &=(\mathbf{J}_{\ell\to z,i}\boldsymbol{\varepsilon}_{\ell,i})^\top
      \mathbf{F}_i(\mathbf{J}_{\ell\to z,i}\boldsymbol{\varepsilon}_{\ell,i}) \nonumber\\
    &\le \lVert \mathbf{F}_i\rVert_2
      \lVert \mathbf{J}_{\ell\to z,i}\boldsymbol{\varepsilon}_{\ell,i}\rVert_2^2 \nonumber\\
    &\le \lVert \mathbf{F}_i\rVert_2
      \lVert \mathbf{J}_{\ell\to z,i}\rVert_2^2
      \lVert\boldsymbol{\varepsilon}_{\ell,i}\rVert_2^2 \nonumber\\
    &\le \kappa_il_{\ell,i}^2
      \lVert\boldsymbol{\varepsilon}_{\ell,i}\rVert_2^2.
    \label{eq:app_quadratic_spectral_steps}
\end{align}
Thus,
\begin{align}
    \Delta l_i^{\mathrm{pred}}
    &\le \frac{1}{2}\kappa_il_{\ell,i}^2
      \lVert\boldsymbol{\varepsilon}_{\ell,i}\rVert_2^2
      +\mathcal O\!\left(\lVert\boldsymbol{\varepsilon}_{\ell,i}\rVert_2^3\right).
    \label{eq:app_spectral_bound}
\end{align}
Defining $c_{\ell,i}=\frac{1}{2}\kappa_il_{\ell,i}^2$ and using
$\lVert\boldsymbol{\varepsilon}_{\ell,i}\rVert_2=\lVert\widetilde{\mathbf{O}}_{\ell,i}-\mathbf{O}_{\ell,i}\rVert_F$, we obtain
\begin{equation}
    \Delta l_i^{\mathrm{pred}}
    \le c_{\ell,i}
      \left\lVert\widetilde{\mathbf{O}}_{\ell,i}-\mathbf{O}_{\ell,i}\right\rVert_F^2
    +\mathcal O\!\left(
      \left\lVert\widetilde{\mathbf{O}}_{\ell,i}-\mathbf{O}_{\ell,i}\right\rVert_F^3
    \right).
    \label{eq:app_prediction_bound}
\end{equation}

\emph{From the layer bound to head errors.}
Let $H$ be the number of heads and
$\Delta\mathbf{O}_{\ell,i,h}$ the matrix of head-$h$ output
perturbations over query block $i$. The projected perturbation satisfies
$\Delta \mathbf{O}_{\ell,i}=\sum_{h=1}^{H}\Delta \mathbf{O}_{\ell,i,h}\mathbf{W}_O^{(h)}$.
Submultiplicativity and the triangle inequality first give
\[
 \|\Delta \mathbf{O}_{\ell,i}\|_F
 \le \sum_h\|\mathbf{W}_O^{(h)}\|_2
              \|\Delta \mathbf{O}_{\ell,i,h}\|_F.
\]
Applying Cauchy--Schwarz to this sum then gives
\begin{equation}
 \|\Delta \mathbf{O}_{\ell,i}\|_F^2
 \le H\sum_h\|\mathbf{W}_O^{(h)}\|_2^2
              \|\Delta \mathbf{O}_{\ell,i,h}\|_F^2.
 \label{eq:app_head_error_bound}
\end{equation}
Substituting this inequality into the leading term of
Eq.~\ref{eq:app_prediction_bound} yields the weighted head-error
bound in Section~\ref{sec:hierarchical_error}. Together, the two bounds link
the final predictive divergence to the reconstruction errors of individual
attention heads and show where head-dependent weights enter the decomposition.

\subsection{Exact Sparse-Attention Output Error}
\label{app:sparse_error_derivation}

We now work within one attention head and retain the original token-level
causal mask inside each block.
The main text writes a single query row with scalar mass $m_{ib}$, conditional
block output $\bar{\mathbf v}_{ib}$, and dense output $\mathbf o_i$. Here a query
block contains several such rows: $\mathbf m_{ib}$ stacks their masses,
$\mathbf D_{ib}=\operatorname{Diag}(\mathbf m_{ib})$ applies them rowwise,
$\mathbf W_{ib}$ stacks the corresponding conditional block outputs, and
$\mathbf O_i^{\mathrm{dense}}$ stacks the dense outputs. Thus each row of
Eq.~\ref{eq:app_exact_error} reduces to the scalar-query identity used in
the main text; the matrix notation introduces no different objective.
For query block $i$, let $\mathbf{A}_{ib}$ denote the dense attention-score
submatrix associated with key block $b$. Define
\begin{equation}
    \mathbf{m}_{ib}=\mathbf{A}_{ib}\mathbf{1},
    \qquad
    \mathbf{D}_{ib}=\operatorname{Diag}(\mathbf{m}_{ib}),
    \qquad
    \mathbf{W}_{ib}=\mathbf{D}_{ib}^{-1}\mathbf{A}_{ib}\mathbf{V}_b.
    \label{eq:app_block_terms}
\end{equation}
Here, $\mathbf{m}_{ib}$ collects the scalar block masses of all query rows in query
block $i$.  For a zero-mass row, we define the corresponding row of $\mathbf{W}_{ib}$ as zero
rather than inverting a zero diagonal entry. Such a row contributes
nothing to $\mathbf{D}_{ib}\mathbf{W}_{ib}$. Since each dense attention
row sums to one, its output decomposes as
\begin{equation}
    \mathbf{O}_i^{\mathrm{dense}}
    =\sum_{b\le i}\mathbf{A}_{ib}\mathbf{V}_b
    =\sum_{b\le i}\mathbf{D}_{ib}\mathbf{W}_{ib}.
    \label{eq:app_dense_decomposition}
\end{equation}

Let $\mathcal T_i$ and $\mathcal R_i$ be the retained and removed block
sets, respectively, and define
\begin{equation}
    \mathbf{D}_{i\mathcal R}
    =\sum_{b\in\mathcal R_i}\mathbf{D}_{ib}.
\end{equation}
Recomputing softmax over the retained tokens divides each query row by its
retained probability mass.  Therefore,
\begin{equation}
    \mathbf{O}_i^{\mathrm{sparse}}(\mathcal T_i)
    =\left(\mathbf{I}-\mathbf{D}_{i\mathcal R}\right)^{-1}
      \sum_{b\in\mathcal T_i}\mathbf{D}_{ib}\mathbf{W}_{ib}.
    \label{eq:app_sparse_output}
\end{equation}
Substituting Eq.~\ref{eq:app_dense_decomposition} gives the exact output
error
\begin{align}
    \mathbf{E}_i(\mathcal T_i)
    &=\mathbf{O}_i^{\mathrm{dense}}
      -\mathbf{O}_i^{\mathrm{sparse}}(\mathcal T_i) \nonumber\\
    &=\left(\mathbf{I}-\mathbf{D}_{i\mathcal R}\right)^{-1}
      \sum_{b\in\mathcal R_i}\mathbf{D}_{ib}
      \left(\mathbf{W}_{ib}-\mathbf{O}_i^{\mathrm{dense}}\right).
    \label{eq:app_exact_error}
\end{align}
The single-token form of Eq.~\ref{eq:app_exact_error} has also been used
to score cached tokens in KV-cache eviction~\citep{goel2025caote,an2026rest}.
For multiple removed blocks, submultiplicativity and the triangle inequality
give
\begin{align}
    \left\|\mathbf{E}_i(\mathcal T_i)\right\|_F
    &\le
    \left\|\left(\mathbf{I}-\mathbf{D}_{i\mathcal R}\right)^{-1}\right\|_2
    \sum_{b\in\mathcal R_i}
    \left\|\mathbf{D}_{ib}
      \left(\mathbf{W}_{ib}-\mathbf{O}_i^{\mathrm{dense}}\right)\right\|_F.
    \label{eq:app_error_bound}
\end{align}
Eq.~\ref{eq:app_error_bound} shows that each removed block affects the
output error through both its attention mass and output deviation. We therefore
rank blocks by the mass-weighted output deviation
$m_{ib}\|\bar{\mathbf{v}}_{ib}-\mathbf{o}_i\|_2$, retaining blocks with larger
estimated effects on the attention output.

\subsection{Approximation for Block Output}
\label{app:pooled_proxy}
POEM approximates each block output by applying a first-order Taylor
expansion at the pooled query. The corresponding diagnostic and complexity
analysis appear in Sections~\ref{app:approximation_diagnostics}
and~\ref{app:runtime_memory}.

The derivation below follows the row-vector convention of the main text.
Accordingly, its QK logit is $\mathbf{q}\mathbf{k}^{\top}$ and its
covariance correction is $\mathbf{q}\mathbf{C}$.

\paragraph{A first-order block-output approximation.}
For a fully visible key block $\mathcal B_b$ of size $n_b$, define
\begin{equation}
 \bar{\mathbf{k}}_b=\frac{1}{n_b}\sum_{t\in\mathcal B_b}\mathbf{k}_t,\qquad
 \bar{\mathbf{v}}_b=\frac{1}{n_b}\sum_{t\in\mathcal B_b}\mathbf{v}_t,\qquad
 \mathbf{C}_b=\frac{1}{n_b}\sum_{t\in\mathcal B_b}
       (\mathbf{k}_t-\bar{\mathbf{k}}_b)^\top(\mathbf{v}_t-\bar{\mathbf{v}}_b).
 \label{eq:app_covariance_definition}
\end{equation}
For the pooled query $\bar{\mathbf{q}}_i$, let
$a_t=\bar{\mathbf{q}}_i(\mathbf{k}_t-\bar{\mathbf{k}}_b)^\top/\sqrt{d_h}$.
Subtracting the common block-mean logit does not change the softmax, and the
centered logits satisfy $\sum_t a_t=0$. Introduce
\begin{equation}
 \alpha_t(u)=\frac{\exp(u a_t)}
 {\sum_{r\in\mathcal B_b}\exp(u a_r)},
 \qquad
 \mathbf{g}(u)=\sum_{t\in\mathcal B_b}\alpha_t(u)\mathbf{v}_t.
\end{equation}
The parameter $u$ interpolates between uniform weights at $u=0$ and the
within-block softmax at $u=1$. At $u=0$,
$\alpha_t(0)=1/n_b$ and $\mathbf{g}(0)=\bar{\mathbf{v}}_b$. Moreover,
\begin{align}
 \alpha_t'(u)
 &=\alpha_t(u)\left(a_t-
   \sum_{r\in\mathcal B_b}\alpha_r(u)a_r\right),\\
 \alpha_t'(0)
 &=\frac{1}{n_b}\left(a_t-\frac{1}{n_b}\sum_r a_r\right)
 =\frac{a_t}{n_b}.
\end{align}
Therefore,
\begin{align}
 \mathbf{g}'(0)
 &=\frac{1}{n_b}\sum_t a_t\mathbf{v}_t \\
 &=\frac{\bar{\mathbf{q}}_i}{\sqrt{d_h}}
   \left[\frac{1}{n_b}\sum_t
   (\mathbf{k}_t-\bar{\mathbf{k}}_b)^\top
   (\mathbf{v}_t-\bar{\mathbf{v}}_b)\right]
 =\frac{\bar{\mathbf{q}}_i}{\sqrt{d_h}}\mathbf{C}_b.
\end{align}
The first-order Taylor expansion
$\mathbf{g}(1)\approx\mathbf{g}(0)+\mathbf{g}'(0)$ gives
\begin{equation}
 \mathbf{g}(1)
 \approx\bar{\mathbf{v}}_b
 +\frac{\bar{\mathbf{q}}_i}{\sqrt{d_h}}\mathbf{C}_b
 =:\widehat{\mathbf{v}}_{ib}.
 \label{eq:app_block_taylor}
\end{equation}
The omitted terms are second order in the centered logits $a_t$. When their
within-block spread is moderate, these terms are typically small and are
therefore omitted; Section~\ref{app:block_output_accuracy} evaluates the
resulting approximation directly.

All configured local blocks, including the diagonal query--key block, are
mandatory and are retained without using this approximation for selection.
The pooled-query estimate is applied only to the remaining optional key
blocks, which precede the local window and are fully visible to every token in
query block $i$. For these optional blocks, all tokens in query block $i$ use
the mean query $\bar{\mathbf{q}}_i$ during scoring instead of their individual
queries. After selection, the final sparse-attention kernel still applies the
exact token-level causal mask. The diagnostic in
Section~\ref{app:block_output_accuracy} measures the effects of query pooling
and the first-order truncation.

\section{Algorithm and Implementation Details}
\label{app:algorithm_details}
This section connects the selection formula to the exact operations used by
Algorithm~\ref{alg:poem}.

\subsection{From Block Estimates to the Selection Score}
\label{app:score_construction}
\paragraph{A coarse global-output reference.}
The pooled QK term in Eq.~\ref{eq:poem_selection_score} is
$z_{ib}^{\mathrm{pool}}=\bar{\mathbf{q}}_i\bar{\mathbf{k}}_b^\top/\sqrt{d_h}$.
Following the main text, the implementation approximates this term with the
antidiagonal scorer; denote its output by
$\widehat z_{ib}^{\mathrm{AD}}\approx z_{ib}^{\mathrm{pool}}$. All optional
blocks evaluated by the scorer are fully visible, fixed-size key blocks;
boundary and local blocks are mandatory and do not participate in the
optional-block ranking. Consequently, the block cardinality is constant over
the scored blocks and no block-dependent cardinality term is needed.
Normalizing the antidiagonal estimates over visible blocks gives the pooled
weights $a_{ib}$ used by Stage~1 and by the coarse output reference:
\begin{equation}
 a_{ib}=\frac{\exp(\widehat z_{ib}^{\mathrm{AD}})}
 {\sum_{c\le i}\exp(\widehat z_{ic}^{\mathrm{AD}})},\qquad
 \widehat{\mathbf{o}}_i=\sum_{b\le i}a_{ib}\bar{\mathbf{v}}_b,\qquad
 r_{ib}^{\mathrm{value}}=\|\widehat{\mathbf{v}}_{ib}-\widehat{\mathbf{o}}_i\|_2.
 \label{eq:app_value_proxy}
\end{equation}
Here $\widehat{\mathbf{o}}_i$ aggregates all visible pooled values and contains
no query-conditioned covariance correction; only the Stage~2 candidate block
outputs include that correction.

Taking the logarithm of the separable importance proxy for positive masses
gives $\log m_{ib}+\log\|\bar{\mathbf{v}}_{ib}-\mathbf{o}_i\|_2$. Substituting
the pooled QK term for the first component and the approximated output
deviation for the second, and introducing $\beta$ yields the score in the main
text. Using the antidiagonal approximation to the pooled QK term, the
implementation evaluates
\begin{equation}
 s_{ib}=\widehat z_{ib}^{\mathrm{AD}}+
 \beta\log r_{ib}^{\mathrm{value}},
 \label{eq:app_implemented_score}
\end{equation}
which is the operational form of Eq.~\ref{eq:poem_selection_score} after
replacing its pooled QK term by the antidiagonal estimate and substituting
$r_{ib}^{\mathrm{value}}$ from Eq.~\ref{eq:app_value_proxy}. Thus the
implementation uses antidiagonal scoring to estimate the first term displayed
in the main text; it does not replace that term by the logarithm of a separate
Stage~1 mass estimator.

\paragraph{Ranking, mass accumulation, and mandatory blocks.}
Stage~1 selects the shortest prefix of blocks in descending $a_{ib}$ order
whose cumulative estimated mass reaches $p_1$ of the total visible mass, then
takes the union with the mandatory sink/local set $\mathcal M_i$. Stage~2
initializes $\mathcal S_i=\mathcal M_i$, credits the masses of mandatory blocks
toward the target, and evaluates and sorts by $s_{ib}$ only the remaining
optional candidates $\mathcal C_i\setminus\mathcal M_i$. It appends them until
\begin{equation}
 \sum_{b\in\mathcal S_i}a_{ib}
 \ge p_{2,g(h)}\sum_{b\in\mathcal C_i}a_{ib}.
 \label{eq:app_estimated_mass_coverage}
\end{equation}
The combined score thus determines only the insertion order; the quantity
accumulated against the threshold is the original pooled weight $a_{ib}$, not
a renormalized softmax of the combined score. Mandatory blocks are never
duplicated; if their mass alone meets the target, no optional block is added.
Block discreteness may still cause overshoot. Note that this criterion covers
pooled attention mass and does not guarantee coverage under exact dense
attention.
The final sparse attention is recomputed from the original token-resolution
Q/K/V tensors.

\section{Experimental Setup}
\label{app:experimental_setup}
This section collects the model, dataset, and controlled-diagnostic protocols;
later diagnostic sections refer back to these populations rather than mixing
their setup with the theoretical derivations.

\subsection{Reconstruction-Error Evaluation Setup}
\label{app:reconstruction_eval_setup}
Table~\ref{tab:metric_errors} compares block-scoring proxies on 
LongBench prompts, each containing exactly 8,192 tokens, using Qwen3-8B.
All methods use the same antidiagonal QK scorer and the pooled weights obtained
by normalizing its scores. The comparison varies only the selection score:
pooled attention mass alone, a STEM-style value-magnitude score with
$\beta=0.2$, and POEM's output-relative score with $\beta=0.2$. This
controlled comparison is distinct from running each method's complete native
implementation.

\paragraph{Fixed selection budget.}
Blocks contain 128 tokens. Each query/head retains at most 16 visible blocks,
corresponding to 2,048 tokens; if fewer than 16 blocks are visible, all are
retained. This is a fixed-count diagnostic, not the top-$p$ protocol used for
the main task results.

\paragraph{Layer and logit metrics.}
For the layer-5, layer-15, and layer-35 columns, only the specified decoder
layer is sparsified and all other layers remain dense. We measure hidden-state
MSE immediately after that layer over all $8,192\times4,096$ elements, then
average the per-text MSE over the 30 paired texts. Layer indices are one-based.
The ``Final logits'' column instead sparsifies all 36 decoder layers and measures
raw-logit MSE after the final normalization and language-model head over all
prompt positions and vocabulary entries. It is neither a KL divergence nor an
isolated attention-head error.

\subsection{Exact-Score Reranking Diagnostic}
\label{app:exact_reranking_setup}
Figure~\ref{fig:metric_value_reranking} uses Llama-3.1-8B-Instruct on
RULER examples. For every evaluated block, we compute the exact dense
attention mass $m_{ib}$, conditional block output
$\bar{\mathbf v}_{ib}$, dense head output $\mathbf o_i$, and exact importance
$I_{ib}=m_{ib}\lVert\bar{\mathbf v}_{ib}-\mathbf o_i\rVert_2$.
Panel (d) compares mass-only selection with exact
output-relative reranking at the same retention setting and reports the
fraction of evaluated heads for which the selected block set changes. This
diagnostic is separate from the benchmark scores in
Table~\ref{tab:main_results}.

\subsection{Ablation Experimental Setup}
\label{app:ablation_setup}
The ablations vary one factor at a time while keeping the remaining model and
selector settings fixed as described in the main text and the reproducibility
configuration below.

\paragraph{Output-deviation weight.}
Figure~\ref{fig:ablation_studies}(a) sweeps
$\beta\in\{0,0.1,0.2,0.3,0.4,0.5\}$ for Llama-3.1-8B, Qwen3-8B, and
Qwen3-14B, and reports the score change relative to $\beta=0$ for each model.

\paragraph{Overall threshold and latency.}
\label{app:score_latency_sweep}
Figure~\ref{fig:ablation_studies}(b) uses Llama-3.1-8B on RULER and sweeps
the overall retention threshold over five levels,
$p\in\{0.75,0.80,0.85,0.90,0.95\}$. At each level, we record the RULER score
and selection latency under the same evaluation setting. The plot reports the
increment associated with raising $p$ to the labeled level, rather than the
absolute result at that level. Specifically, the point labeled $p=0.80$
reports the score difference between $p=0.80$ and $p=0.75$, together with the
corresponding latency difference; the points labeled $0.85$, $0.90$, and
$0.95$ are defined analogously relative to their immediately preceding
levels. The five evaluated thresholds therefore produce four plotted
transitions. The $p=0.75$ result serves as the reference for the first
transition and is not shown as a separate point.

\paragraph{Stage~1 threshold.}
Figure~\ref{fig:ablation_studies}(c) uses Llama-3.1-8B on PG19, fixes the
overall target at $p=0.90$, and sweeps
$p_1\in\{0.90,0.91,0.92,0.93,0.94,0.95\}$. All other settings are fixed, and
each point reports word perplexity under the corresponding candidate-pool
threshold.

\paragraph{Scoring proxy and density.}
Figure~\ref{fig:ablation_studies}(d) uses Llama-3.1-8B on PG19 and compares
mass-only, STEM, and POEM scoring at matched block densities. Each point
reports word perplexity, with the inputs and density fixed across the three
selection scores.

\paragraph{Component ablation.}
\label{app:pg19_component_ablation}
Table~\ref{tab:cov_deviation_ablation} evaluates Qwen3-8B on 20 books from
the PG19 test split, using 32K-token inputs and a fixed attention density of
approximately $0.1$. The same books and density are used for POEM, the
variant without covariance correction, and the variant without the
output-relative reference. Word perplexity is computed over this fixed
20-book evaluation population.

\subsection{Models, Datasets, and Evaluation Scope}
\label{app:models_datasets}
\paragraph{Backbones and context lengths.}
We evaluate Llama-3.1-8B~\citep{grattafiori2024llama} and
Qwen3-8B/14B~\citep{yang2025qwen3}. Model parameters remain frozen. The
long-context Qwen runs use YaRN~\citep{peng2026yarnefficientcontextwindow} to extend the 32K context
window to 128K tokens; this setting should not be inferred for every 8K
diagnostic.

\paragraph{Datasets and reporting.}
LongBench~\citep{bai2024longbench} contains 21 tasks in six categories; we use
each prescribed metric and report the unweighted macro average of category
scores. RULER~\citep{hsieh2024ruler} is reported at 4K, 8K, 16K, 32K, 64K,
and 128K and is averaged uniformly over lengths. PG19~\citep{raecompressive2019}
supplies long-form text for perplexity ablations and selected block
diagnostics. Task evaluation uses lm-evaluation-harness~\citep{eval-harness};
component diagnostics use the explicit samples, precision, thresholds, and
aggregation rules stated with each diagnostic and are not pooled into the
task benchmarks.

\subsection{Reproducibility Configuration}
All models are inference-only and retain their released weights. Experiments
use PyTorch on an NVIDIA A100, 128-token query/key blocks, antidiagonal
scoring with $B_1=16$ in Stage~1 and $B_2=8$ in Stage~2, and
FlashAttention-2 for the final token-resolution attention.
Llama-3.1-8B uses four sink and four local blocks; Qwen3-8B/14B use one of
each. Stage~1 uses $p_1=0.94$. The output-deviation coefficient is $0.3$ for
Llama-3.1-8B and Qwen3-14B and $0.2$ for Qwen3-8B. Before calibration,
$p_2=0.90/0.94$; calibrated groups use $0.81/0.94$. The logarithm floor is
$10^{-6}$.

The diagnostic subsections additionally state model trajectory, activation
precision, prompt length, sample count, layer indexing, candidate definition,
and aggregation unit. A reproduction artifact should preserve the ordered
example identifiers and tokenized lengths for every diagnostic and for the
two calibration manifests in Section~\ref{app:threshold_calibration}; this
prevents nominally identical datasets from silently producing different
subsets after tokenization or filtering.

\section{Approximation and Preselection Diagnostics}
\label{app:approximation_diagnostics}
We first test whether the two output estimates are faithful to their distinct
targets, and then test the operational question for preselection: whether the
mass-only candidate pool excludes blocks that an all-visible output-relative
selector would retain. The block-output and global-output plots use different
units and populations and are therefore complementary diagnostics, not a
head-to-head comparison.

\begin{figure}[H]
 \centering
 \includegraphics[width=\textwidth]{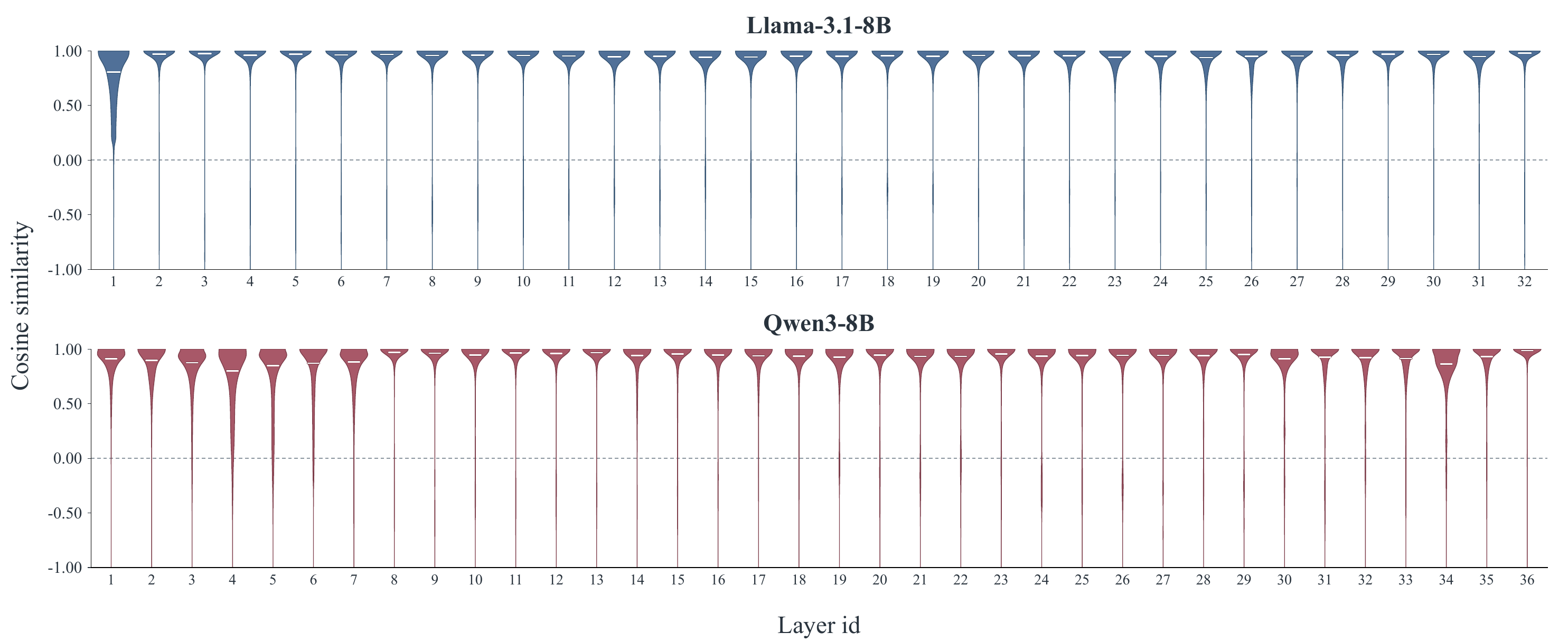}
 \caption{Cosine similarity between the first-order estimate
 $\widehat{\mathbf{v}}_{ib}$ and the exact token-level causal block output
 $\mathbf{v}_{ib}^{\mathrm{exact}}$. White bars mark weighted medians.}
 \label{fig:block_output_estimation_accuracy}
\end{figure}

\subsection{Block-Output Approximation}
\paragraph{Block-output diagnostic: reference and evaluation population.}
\label{app:block_output_accuracy}
We use 30 shared LongBench text identities for Llama-3.1-8B and
Qwen3-8B, tokenize them separately for each model, and retain 8,192
tokens without a chat template. Activations follow the dense BF16 model
trajectory. All layers, 32 query heads, and 64 query blocks are evaluated.
Blocks contain 128 tokens. Within each layer, every text/head/query-block row
has equal total weight, divided equally among its evaluated key blocks. Layers
are indexed from one in the figure.

For query block $\mathcal Q_i$ and key block $\mathcal B_b$, the reference is
\begin{equation}
 \mathbf{v}_{ib}^{\mathrm{exact}}
 =\frac{1}{|\mathcal Q_i|}\sum_{t\in\mathcal Q_i}
 \frac{\sum_{u\in\mathcal B_b,\,u\le t}
       \exp(\mathbf{q}_t\mathbf{k}_u^\top/\sqrt{d_h})\mathbf{v}_u}
      {\sum_{u\in\mathcal B_b,\,u\le t}
       \exp(\mathbf{q}_t\mathbf{k}_u^\top/\sqrt{d_h})}.
 \label{eq:app_exact_block_output_reference}
\end{equation}
We compare the reference with
$\widehat{\mathbf{v}}_{ib}=\bar{\mathbf{v}}_b+
\bar{\mathbf{q}}_i\mathbf{C}_b/\sqrt{d_h}$ using cosine similarity:
\begin{equation}
 c_{ib}=\frac{\widehat{\mathbf{v}}_{ib}(\mathbf{v}_{ib}^{\mathrm{exact}})^{\top}}
                  {\|\widehat{\mathbf{v}}_{ib}\|_2\|\mathbf{v}_{ib}^{\mathrm{exact}}\|_2}
 \label{eq:app_block_signed_cosine}
\end{equation}
for each text, layer, head, and query--key block pair. Equal row weighting
prevents query blocks with more evaluated key blocks from dominating the
distribution.

\paragraph{Block-output cosine similarity.}
Figure~\ref{fig:block_output_estimation_accuracy} shows high cosine similarity
for most evaluated blocks. The weighted median exceeds $0.9$
in 31 of 32 Llama layers and 29 of 36 Qwen layers. The median is lower in
Llama's first layer ($0.805$), several early Qwen layers, and Qwen layer
34. Thin lower tails also indicate poorly aligned pairs in otherwise
well-aligned layers. These results support the first-order estimate as a
useful block-level signal while identifying layer-dependent limitations.
Cosine similarity measures direction, not vector-magnitude error;
it does not by itself bound the error in the block-output norm or guarantee
that approximate scores preserve the exact block ranking.

\subsection{Accuracy of the Global Output Estimate}
\label{app:output_estimation_accuracy}
The global reference $\widehat{\mathbf{o}}_i$ anchors the output deviation used
for block selection. To examine how much information this inexpensive
estimate retains about the final attention output, we compare it with the
dense attention output before the multi-head output projection.

\begin{figure}[!htb]
 \centering
 \includegraphics[width=\textwidth]{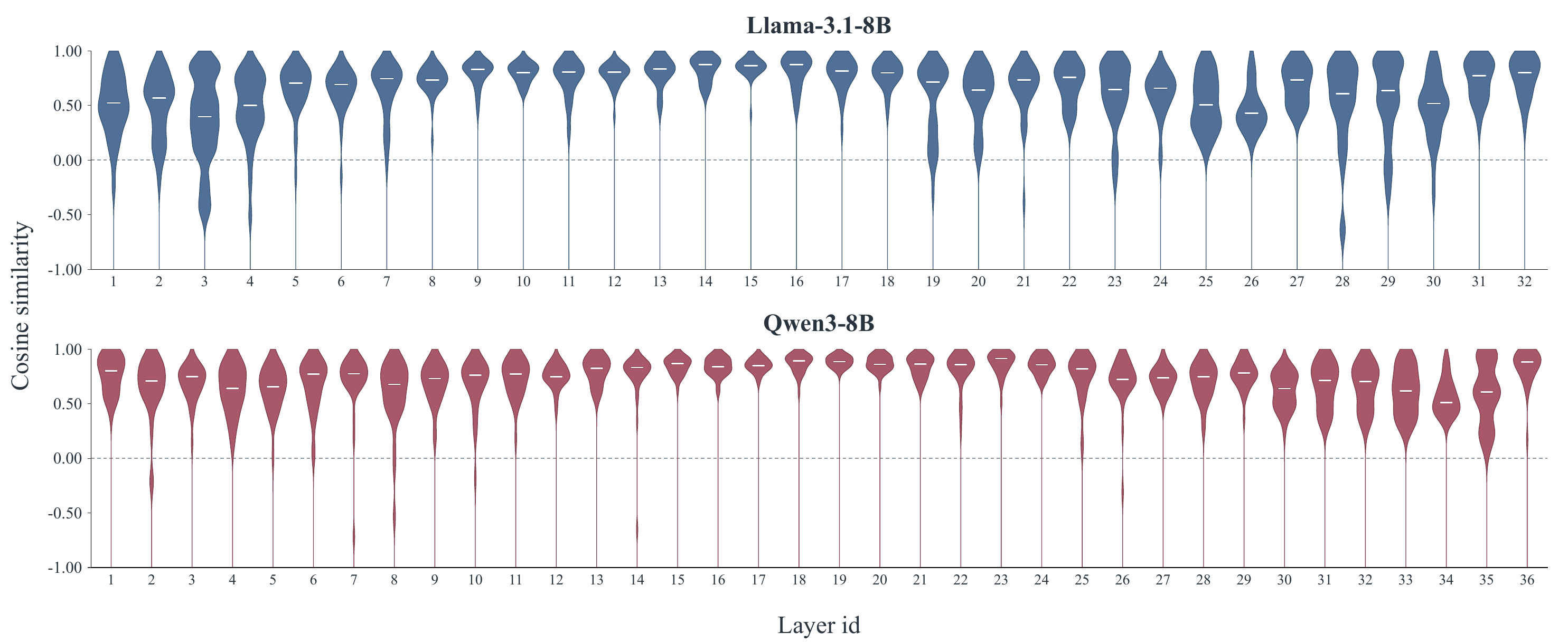}
 \caption{Head-level cosine similarity between the Stage~1 global-output
 estimate and the dense attention output across layers.}
 \label{fig:output_estimation_accuracy}
 \label{fig:output_cosine}
\end{figure}

\paragraph{Reference and aggregation.}
Figure~\ref{fig:output_estimation_accuracy} evaluates the raw pooled
reference $\widehat{\mathbf{o}}_i$ on 30 LongBench texts of 8,192 tokens and
128-token blocks. The output is estimated using the Stage~1 procedure with
$p_1=0.94$. Let $\mathcal X$ contain all text--block pairs. For each layer
$\ell$ and query head $h$, we compute one cosine similarity by summing the
dot products and squared norms over all pairs in $\mathcal X$ before taking
their ratio:
\begin{equation}
 c_{\ell h}^{\mathrm{global}}
 =\frac{\sum_{x\in\mathcal X}\widehat{\mathbf{o}}_{\ell hx}^{\top}
 \mathbf{o}_{\ell hx}^{\mathrm{dense}}}
 {\sqrt{\sum_{x\in\mathcal X}\|\widehat{\mathbf{o}}_{\ell hx}\|_2^2}
 \sqrt{\sum_{x\in\mathcal X}\|\mathbf{o}_{\ell hx}^{\mathrm{dense}}\|_2^2}}.
 \label{eq:app_global_signed_cosine}
\end{equation}
Thus, each layer contributes 32 head-level cosine values to the violin plot.

\paragraph{Results and interpretation.}
Every layer has a positive median pooled cosine. The layer medians range
from $0.399$ to $0.873$ for Llama and from $0.510$ to $0.914$ for Qwen,
with stronger alignment across many middle layers. Several layers have
a broad spread across heads, and some heads have negative pooled cosine.
These results show that the low-cost estimate captures a meaningful portion
of the dense output information and can serve as a coarse prior for block
selection. However, the variation across layers and heads also shows that
this prior is only approximate and provides limited output information in
the less-aligned cases.

\subsection{Stage~1 Candidate-Pool Coverage}
\label{app:stage1_value_effect}
Stage~1 selects candidates using attention mass alone, whereas Stage~2
combines routing relevance with value-dependent output deviation.
We first answer the operational question---whether preselection excludes
blocks that the combined score would retain---and then use the component
distributions to explain the observed coverage.

\paragraph{Diagnostic setup.}
All four figures in this section use Llama-3.1-8B at layers 5, 15, 25,
and 31 (zero-based), evaluated on two fixed PG19 texts at nested input
lengths of 16K, 32K, 48K, and 64K. The activations follow the same dense
BF16 trajectory. Each layer/length combination contains 512 query rows:
two texts, 32 heads, and eight sampled query-block rows per text and
head. We retain the diagnostic's native antidiagonal QK scores, their
normalized pooled weights, covariance-corrected value deviations, and frozen
output-reference calibration. The distribution plots apply no additional
per-row, per-head, or per-layer normalization to these recorded deviations.
Keeping the sampled inputs and layer indices aligned makes it possible
to compare the mass profiles, deviation distributions, and selection
coverage within this diagnostic.

\begin{figure}[!htb]
  \centering
  \includegraphics[width=\textwidth]{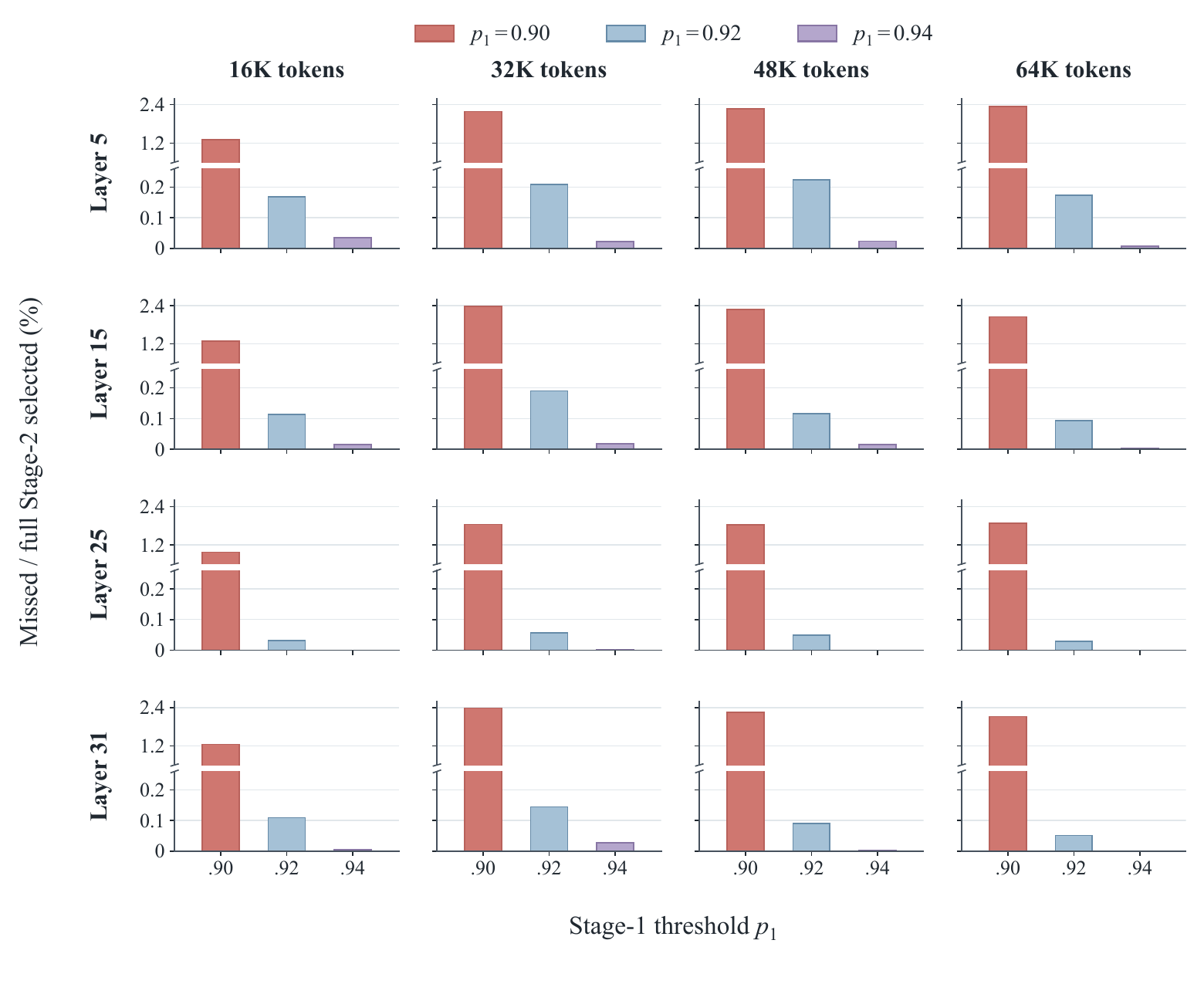}
  \caption{Layerwise coverage of value-aware selections by mass-only
  preselection for $p_1\in\{0.90,0.92,0.94\}$. The missed fraction is the
  pooled count of all-visible reference-selected blocks outside the Stage~1
  pool divided by the pooled reference-selected count. At $p_1=0.94$, the
  missed fraction is at most $0.035\%$ across all 16 measured layer/length
  combinations.}
  \label{fig:stage1_missed_selections}
\end{figure}

\paragraph{Core coverage result.}
Figure~\ref{fig:stage1_missed_selections} uses an all-visible Stage~2
reference: output-relative selection is applied to all causally visible
blocks with a raw mass target of $0.90$ and $\beta=0.6$. The reference is
fixed while $p_1$ changes. With $\mathcal S_{i,h}^{\mathrm{full}}$ denoting
the reference selection and $\mathcal C_{i,h}(p_1)$ the mass-only pool,
\begin{equation}
 r_{\mathrm{miss}}(p_1)=
 \frac{\sum_{i,h}|\mathcal S_{i,h}^{\mathrm{full}}\setminus
 \mathcal C_{i,h}(p_1)|}
 {\sum_{i,h}|\mathcal S_{i,h}^{\mathrm{full}}|}.
 \label{eq:app_stage1_miss_rate}
\end{equation}
The missed fraction is $0.973$--$2.383\%$ at $p_1=0.90$,
$0.029$--$0.224\%$ at $p_1=0.92$, and at most $0.035\%$ at $p_1=0.94$.
Thus the deployed conservative pool covers at least $99.96\%$ of the
reference-selected blocks in every measured layer/length combination.

\begin{figure}[!htb]
  \centering
  \includegraphics[width=\textwidth]{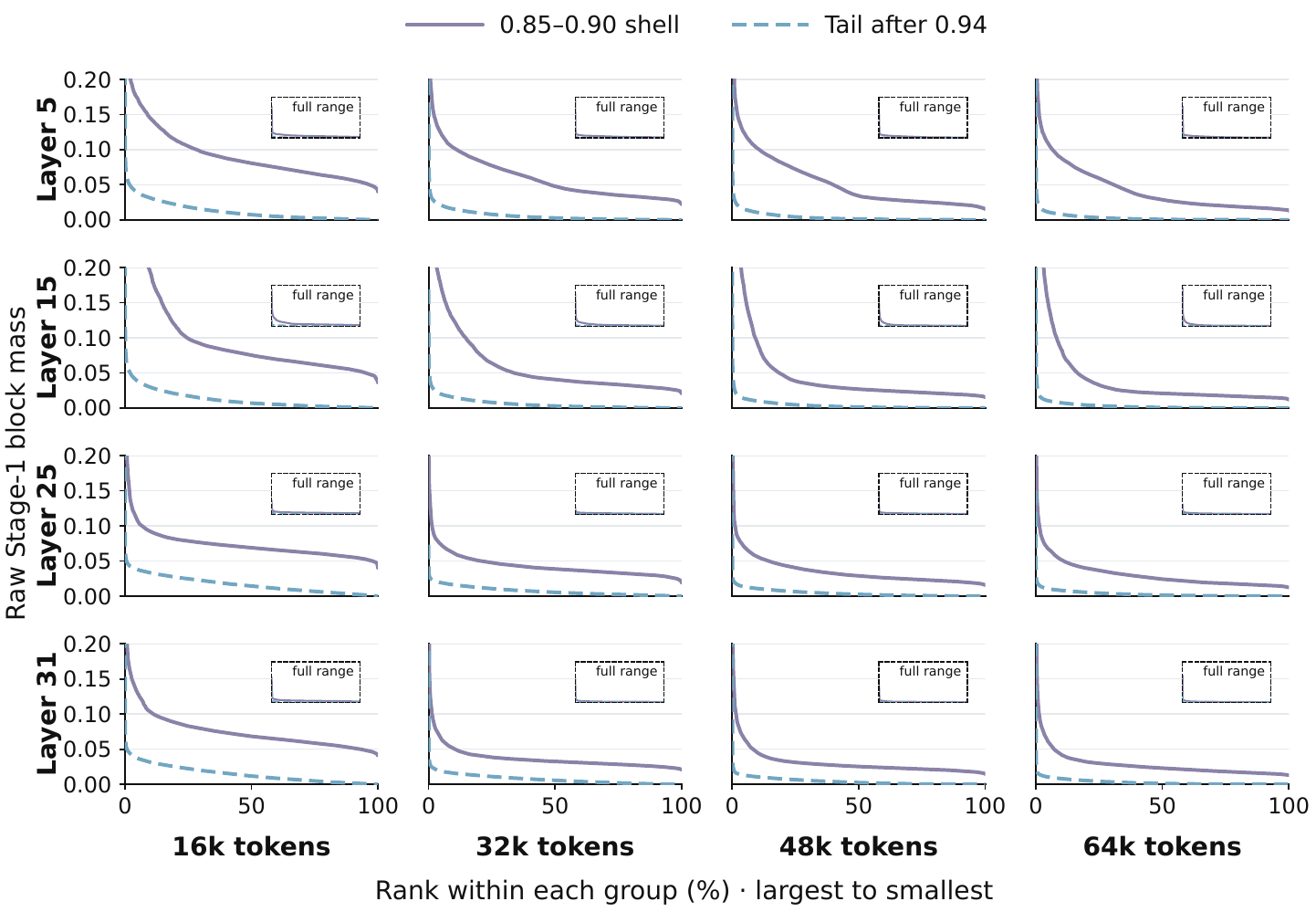}
  \caption{Attention-mass profiles for the $0.85$--$0.90$ cumulative-mass
  shell and the tail beyond $0.94$, sorted independently within each group.
  The tail has substantially lower raw Stage~1 block mass over most ranks
  across all sampled layers and context lengths.}
  \label{fig:stage1_mass_profiles}
\end{figure}

\begin{figure}[!htb]
  \centering
  \includegraphics[width=\textwidth]{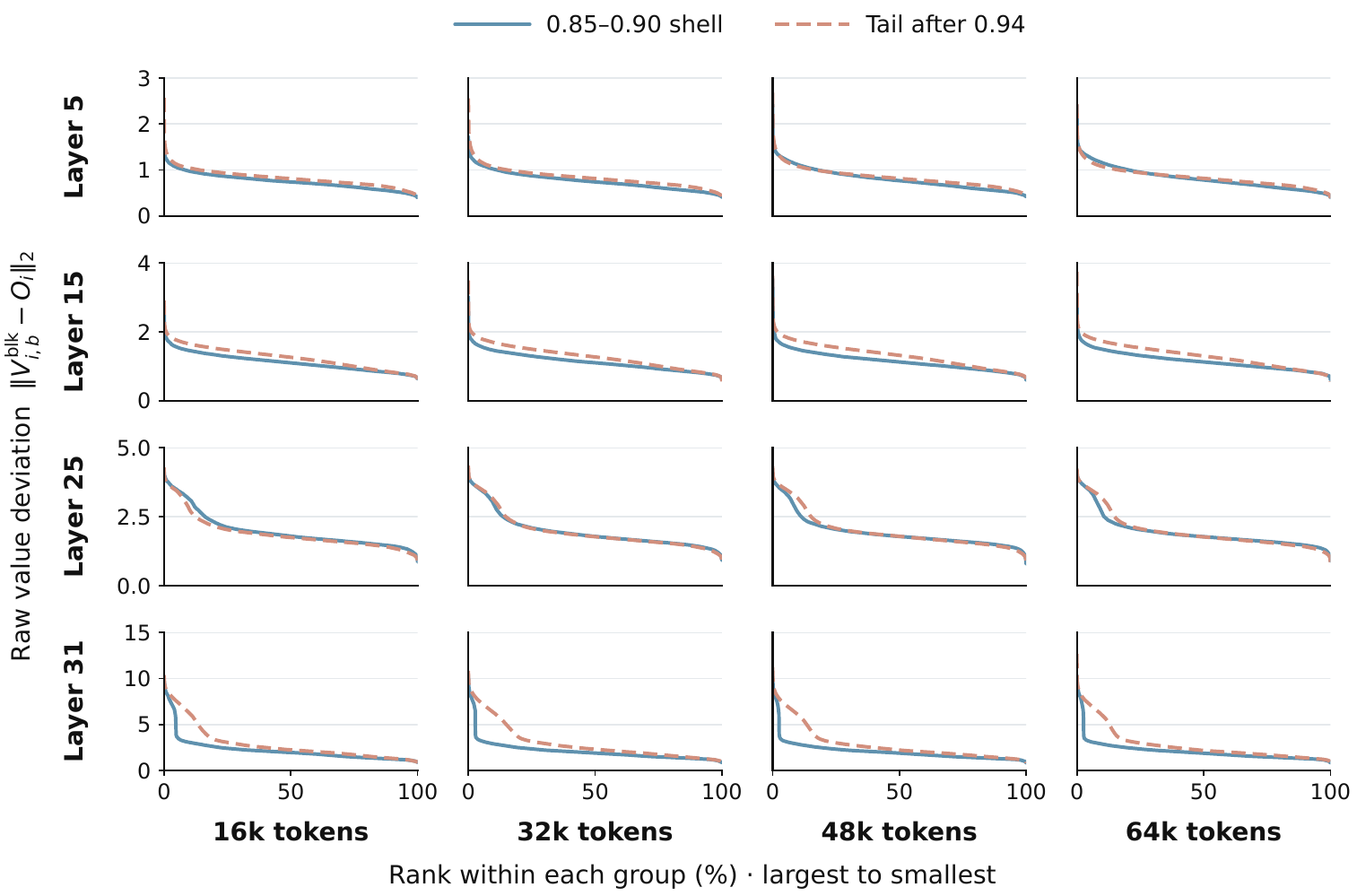}
  \caption{Output-deviation profiles for the $0.85$--$0.90$ cumulative-mass
  shell and the tail beyond $0.94$, sorted independently within each group.
  The profiles largely overlap except for the upper-ranked tail in later
  layers, most notably layer 31.}
  \label{fig:stage1_value_profiles}
\end{figure}

\paragraph{Attention mass versus value deviation.}
Figures~\ref{fig:stage1_mass_profiles}
and~\ref{fig:stage1_value_profiles} compare two groups defined by cumulative
attention mass: the shell between $0.85$ and $0.90$, and the tail beyond
$0.94$.
The plots cover layers 5, 15, 25, and 31 and input lengths from 16K to 64K (zero-based layer indices).

\paragraph{Attention-mass separation.}
Figure~\ref{fig:stage1_mass_profiles} shows a clear and persistent gap
between the two groups. Across every sampled layer and context length, the
$0.85$--$0.90$ shell retains markedly larger raw Stage~1 block mass over
almost the entire within-group rank range. By contrast, the tail beyond
$0.94$ drops rapidly and approaches zero for a large fraction of its
blocks. The same ordering appears from 16K through 64K and from the early
layer 5 to the late layer 31, although the precise curve shape changes with
depth. This separation means that a tail block begins with a substantial
QK-score disadvantage before the value-aware term is applied. Consequently,
only tail blocks with unusually large output deviation can plausibly
overtake candidates already included in the conservative mass pool.

\paragraph{Output-deviation overlap.}
Figure~\ref{fig:stage1_value_profiles} presents a different pattern. The
ranked output-deviation curves of the shell and tail largely overlap in
layers 5 and 25, while the tail is only moderately higher through much of
layer 15. The strongest difference occurs in layer 31, where the upper
portion of the tail has larger deviations than the corresponding shell
profile. Even there, the gap contracts toward the lower-ranked portion of
the distributions, and the two groups remain on a comparable scale over
most ranks. Thus, low attention mass does not imply uniformly negligible
value deviation, but neither does the tail exhibit a broad value advantage
that would generally compensate for the mass gap in
Figure~\ref{fig:stage1_mass_profiles}.

\paragraph{Joint interpretation.}
Together, the two figures motivate the asymmetric two-stage design. Stage~1
uses attention mass to remove a region whose scores are consistently weak,
whereas Stage~2 preserves the value term for finer discrimination inside a
conservative candidate pool. The figures are distributional rather than
paired blockwise comparisons: ranks are computed independently within each
group and component, so points at the same horizontal position need not
represent the same block. The profiles therefore explain the empirical
coverage trend without asserting that mass alone determines the final
selection.

For positive deviations, a discarded block $b$ can outrank a candidate
$c$ under Eq.~\ref{eq:poem_selection_score} only if
\begin{equation}
    \beta\log\frac{r_{ib}^{\mathrm{value}}}{r_{ic}^{\mathrm{value}}}
    > \widehat z_{ic}^{\mathrm{AD}}-\widehat z_{ib}^{\mathrm{AD}}.
    \label{eq:app_value_rank_reversal}
\end{equation}
Here $\widehat z_{ib}^{\mathrm{AD}}$ is the antidiagonal estimate of the pooled
QK component in the combined log score.
For $\beta>0$, a large routing-score gap requires a correspondingly large
value-deviation ratio. The observed profiles support using the value term
to refine a conservative candidate pool rather than scoring every
low-mass block. This explains why the conservative pool has high measured
coverage without claiming that attention mass alone is sufficient for the
final ranking.

\begin{figure}[!htb]
    \centering
    \includegraphics[width=\textwidth]{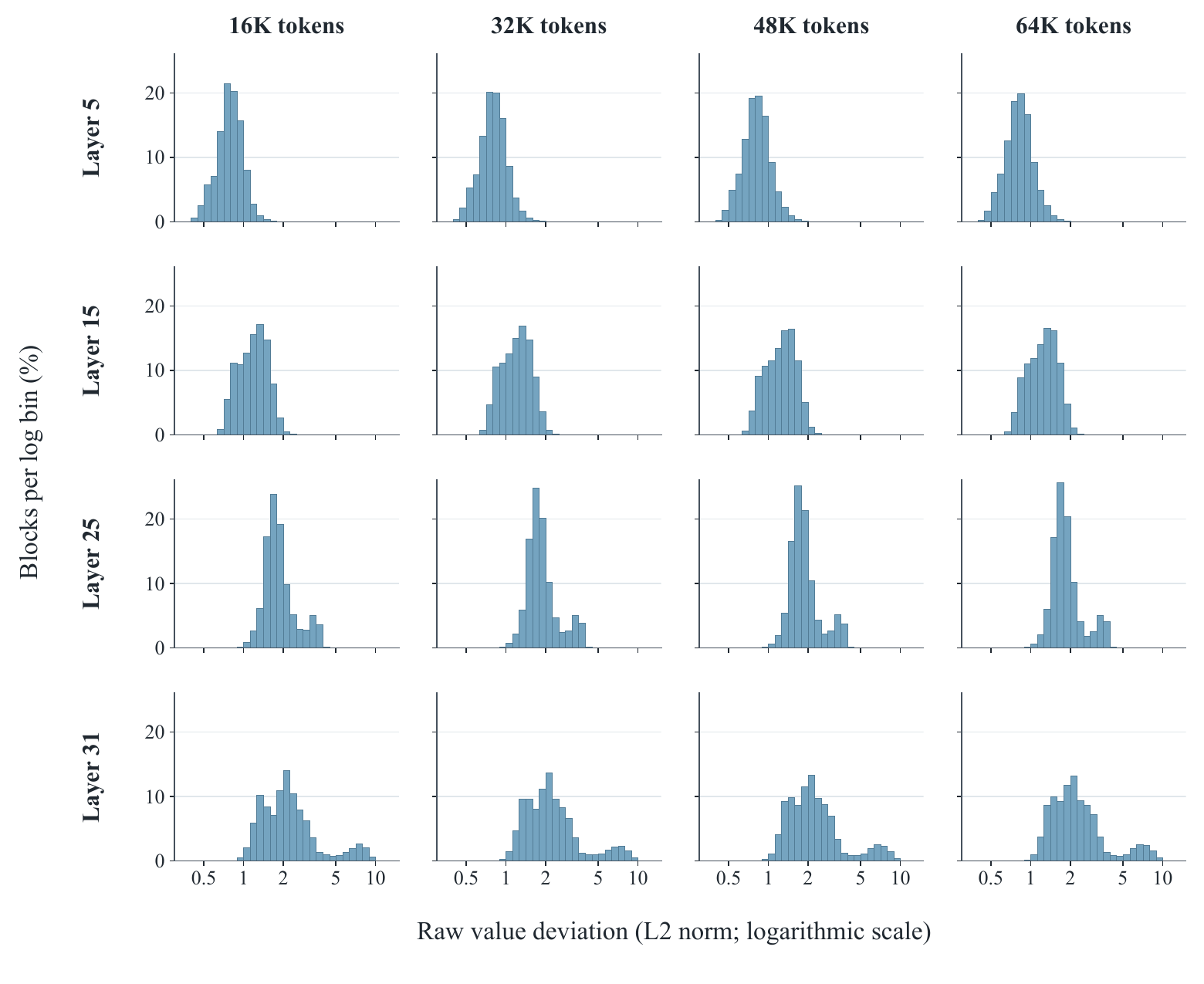}
    \caption{Layerwise distributions of raw value-deviation norms over
    all visible blocks. Rows show Llama-3.1-8B layers 5, 15, 25, and 31
    (zero-based); columns show four input lengths. Each panel pools
    two PG19 texts, 32 heads, and eight query rows per text and head,
    with equal weight per visible block observation. The horizontal
    axis is logarithmic, and bins have width $0.05$ in $\log_{10}$
    deviation. All panels share the same axes, without additional
    per-head or per-row normalization. Layer 31 has a broader right
    tail, while each layer's central distribution varies modestly
    across lengths.}
    \label{fig:stage1_value_histograms}
\end{figure}

\paragraph{Layerwise distribution of value deviation.}
Figure~\ref{fig:stage1_value_histograms} separates the all-visible
deviation distribution by layer and input length. Each panel pools the
visible block observations for its 512 query rows, with equal weight per
block observation. The 16 panels contain 2,295,552 observations in total.
The vertical axis reports the percentage of observations in each
$0.05$-wide $\log_{10}$ bin. Common axes retain the original norm scale
and all observed values, so both location and spread can be compared
across panels. The distribution includes retained and discarded blocks;
it is not conditioned on either stage's final selection.

The typical deviation increases across the four sampled depths. Across
the four lengths, the median is $0.787$--$0.817$ in layer 5,
$1.202$--$1.267$ in layer 15, $1.765$--$1.779$ in layer 25, and
$2.076$--$2.085$ in layer 31. Within each layer, the largest median is
less than $6\%$ above the smallest across these lengths. The late layer
also has a broader right tail: its 95th percentile is
$6.864$--$7.180$, compared with at most $3.501$ in the other three
layers. Nevertheless, deviations above 10 account for only
$0.037$--$0.050\%$ of layer-31 observations and do not occur in the
other plotted layers.

These layerwise views explain why a pooled histogram can obscure
meaningful depth-dependent differences. They also clarify the role of
the value term: most blocks have moderate deviations, while a small
tail can receive substantially higher scores. Whether such a block
should survive depends jointly on its deviation and attention mass,
as expressed by Eq.~\ref{eq:app_value_rank_reversal}. A marginal
deviation histogram alone cannot establish that preselection preserves
the desired blocks; the following coverage test evaluates that question
using the joint score.

The threshold effect is consistent across the four depths, including
layer 31 with its broader deviation tail. The length dependence is not
strictly monotone: for example, at $p_1=0.90$, layer 15 reaches
$2.378\%$ at 32K and decreases to $2.053\%$ at 64K. Hence the observed
coverage is better summarized by its range across layers and lengths
than by assuming that longer inputs necessarily cause more omissions.

Together, the profiles and coverage measurements support using a
conservative mass-only pool before applying the more expensive
output-relative score. The profiles identify the large mass gap that a
tail block must overcome, while the coverage test directly counts the
exceptions under the combined score. The diagnostic uses $\beta=0.6$,
rather than the model-specific coefficients of the main benchmark.
Its two-text coverage result does not guarantee zero omissions on other
inputs, nor does a small missed-block fraction alone bound output error:
missed blocks can differ in their individual contributions.

\section{Runtime and Memory Analysis}
\label{app:runtime_memory}

\subsection{Block-Output Estimator FLOPs}
The relevant baseline is the exact token-level block output in
Eq.~\ref{eq:app_exact_block_output_reference}. Let
\begin{equation}
 n_{ib}=\sum_{t\in\mathcal Q_i}|\{u\in\mathcal B_b:u\le t\}|
\end{equation}
be the number of visible query--key pairs. Counting a multiply--add as two
FLOPs, exact QK products and value accumulation cost
\begin{equation}
 f_{\mathrm{exact}}(i,b)=4n_{ib}d_h,
 \label{eq:app_exact_block_flops}
\end{equation}
where $n_{ib}=b_qb_k$ for a fully visible block. A cached covariance
descriptor requires one matrix--vector product and one vector addition,
\begin{equation}
 f_{\mathrm{Taylor}}=2d_h^2+d_h.
 \label{eq:app_taylor_flops}
\end{equation}
For a fully visible block, the fraction of per-candidate arithmetic retained
by the estimator is therefore
\begin{equation}
 \frac{f_{\mathrm{Taylor}}}{f_{\mathrm{exact}}}
 =\frac{2d_h+1}{4b_qb_k},
 \qquad
 \text{saving}=1-\frac{2d_h+1}{4b_qb_k}.
 \label{eq:app_taylor_flop_reduction}
\end{equation}
When $b_q=b_k=d_h=d$, exact block computation is
$\mathcal O(d^3)$ whereas the estimate is $\mathcal O(d^2)$ per candidate.
At $d=128$, Eq.~\ref{eq:app_taylor_flop_reduction} gives a $99.61\%$
reduction in per-candidate arithmetic.

The covariance costs
$2b_kd_h^2+\mathcal O(b_kd_h+d_h^2)$ to construct, but it is computed once
per KV-head/key-block descriptor and reused across the corresponding query
heads and query blocks. Query pooling likewise costs $2b_qd_h$ once per query
block. If $P$ candidate pairs reuse $U$ unique descriptors across $Q$ query
blocks, the estimator requires
\begin{equation}
 P(2d_h^2+d_h)
 +U\!\left[2b_kd_h^2+\mathcal O(b_kd_h+d_h^2)\right]
 +2Qb_qd_h
 \label{eq:app_taylor_total_flops}
\end{equation}
FLOPs, compared with $4d_h\sum_{(i,b)}n_{ib}$ for exact token-level block
outputs. The descriptor and pooling terms are therefore amortized as their
reuse increases. For a causal diagonal block,
Eq.~\ref{eq:app_exact_block_flops} uses the actual smaller $n_{ib}$, so
its saving is lower than the full-block expression. These counts exclude
memory traffic, kernel launches, Stage~1, and the final sparse attention.
End-to-end measurements appear in
Figures~\ref{fig:efficiency_breakdown} and~\ref{fig:poem_stage_timing}.

\subsection{Memory Overhead of Q/K Copies and Covariance}
\label{app:memory_overhead}
We first calculate the storage required by the auxiliary tensors used for
low-precision scoring and block-output approximation. These tensors consist
of low-precision Q/K copies and a blockwise key--value covariance cache. In
practice, our implementation stores the Q/K copies in INT8 and the covariance
cache in BF16. We then measure the actual instantaneous memory overhead of
the INT8 Q/K copies and covariance cache.

\paragraph{Storage calculation.}
For batch size one, sequence length $n$ divisible by block size $b=128$, head
dimension $d_h$, $h_q$ query heads, and $h_{kv}$ KV heads, let
$b_{\mathrm{tot}}=\lceil n/b\rceil$. An INT8 entry occupies one byte, and a BF16
covariance entry occupies two bytes. The buffer sizes are
\begin{equation}
 m_{\mathrm{copies}}=n d_h(h_q+h_{kv})\ \text{bytes},\qquad
 m_{\mathrm{cov}}=2h_{kv}b_{\mathrm{tot}} d_h^2\ \text{bytes}.
 \label{eq:poem_auxiliary_memory}
\end{equation}
The covariance cache is shared by query heads using the same KV head.
At 128K with $d_h=128$, $h_q=32$, and $h_{kv}=8$, the Q and K copies
require 512 and 128 MiB, respectively, and the covariance cache requires
256 MiB.

\paragraph{Verification with real model activations.}
We replay saved dense-trajectory Q/K/V activations from Qwen3-8B
(YaRN $\times4$, layer index 18) for four fixed inputs: two PG19 texts
and two InfiniteBench inputs (code debugging and key--value retrieval).
The actual INT8 quantization and BF16 covariance routines are executed
on an NVIDIA A100 80GB at 16K, 32K, 64K, and 128K, with batch size one.
After warm-up, we measure the actual instantaneous memory increase produced
by the INT8 Q/K copies and covariance cache during complete POEM execution.

\begin{table}[!htbp]
 \centering
 \caption{Predicted and measured auxiliary tensor storage (MiB).
 The actual INT8 copies and BF16 covariance cache match their shape-based
 predictions on all four inputs at each length. Q/K measurements also
 include the small FP32 quantization-scale tensors.}
 \label{tab:poem_memory_estimate}
 \small
 \setlength{\tabcolsep}{7pt}
 \begin{tabular}{lrrrr}
  \toprule
  & \multicolumn{2}{c}{Q/K copies} & \multicolumn{2}{c}{Covariance} \\
  \cmidrule(lr){2-3}\cmidrule(lr){4-5}
  Length & Predicted payload & Measured with scales & Predicted & Measured \\
  \midrule
  16K  & 80  & 80.020  & 32  & 32 \\
  32K  & 160 & 160.039 & 64  & 64 \\
  64K  & 320 & 320.078 & 128 & 128 \\
  128K & 640 & 640.156 & 256 & 256 \\
  \bottomrule
 \end{tabular}
\end{table}

Table~\ref{tab:poem_memory_estimate} confirms that these buffers occupy
the predicted storage in the implementation. The measured Q/K difference
is exactly the FP32 scale storage, $4(h_q+h_{kv})b_{\mathrm{tot}}$ bytes; at 128K it is
0.15625 MiB. All 48 component checks (Q copy, K copy, and covariance
across 16 input/length combinations) match the tensor payload plus its
returned metadata exactly. Instrumented and uninstrumented complete
POEM calls produce bitwise-identical attention outputs on all 16 cases.

Both terms grow linearly with $n$ for fixed block size and head geometry.
Their payload sizes sum to 896 MiB at 128K, but this sum is not a claim
about the complete model's peak GPU memory. The INT8 copies are temporary
Stage~1 tensors, while covariance is constructed for Stage~2; their
lifetimes need not overlap. This accounting is limited to the two
buffer families and excludes the original Q/K/V activations and other
attention workspaces.

\paragraph{End-to-end peak GPU memory.}
We additionally record the peak allocated GPU memory during full-model
prefilling. Table~\ref{tab:poem_model_peak_memory} shows that the runtime
memory peaks of POEM and dense attention are nearly identical. Even at
128K tokens, POEM increases the peak by only 2.005 MiB.

\begin{table}[H]
 \centering
 \caption{End-to-end peak allocated GPU memory during full-model
 prefilling. The increase is computed from the unrounded measurements.}
 \label{tab:poem_model_peak_memory}
 \small
 \setlength{\tabcolsep}{12pt}
 \begin{tabular}{lrrr}
  \toprule
  Length & Dense (GiB) & POEM (GiB) & Increase (MiB) \\
  \midrule
  16K  & 19.147 & 19.147 & 0.033 \\
  32K  & 23.030 & 23.031 & 0.127 \\
  64K  & 30.796 & 30.797 & 0.503 \\
  128K & 46.329 & 46.331 & 2.005 \\
  \bottomrule
 \end{tabular}
\end{table}

The model-level peak is determined by the allocation of MLP activations.
When this peak is reached, the temporary low-precision Q/K copies and the
covariance cache used by POEM have already been released. These buffers
therefore do not overlap with the dominant MLP allocation, explaining why
their standalone storage does not translate into a comparable increase in
the end-to-end peak memory.

\section{Offline Threshold Calibration}
\label{app:threshold_calibration}
Algorithm~\ref{alg:calibration} expands the error-ranked batch-bisection
procedure in Section~\ref{sec:headwise_budget}. A group is a
layer-specific set of query heads sharing KV projections. Its members
share one threshold. To keep calibration notation consistent with the main
configuration, let $\theta_g=p_1p_{2,g}$ denote the group's overall estimated
mass target; the candidate-relative threshold used by Algorithm~\ref{alg:poem}
is $p_{2,g}=\theta_g/p_1$. Groups are ordered by increasing error in
a single-group perturbation diagnostic, with all other groups kept at
the reference setting. This ordering prioritizes groups expected to
tolerate a smaller budget.

\paragraph{Data partition.}
Calibration uses two fixed, disjoint manifests: $\mathcal D_{\mathrm{rank}}$
for the dense-logit cache and single-group KL probes, and
$\mathcal D_{\mathrm{search}}$ for accepting or rejecting threshold changes.
The benchmark test instances reported in Section~\ref{sec:experiment} belong
to neither manifest and are evaluated only after thresholds are frozen. The
example identifiers, tokenized lengths, sampled prediction positions, and
random seed must be stored with the released configuration so that the split
can be reconstructed exactly.

\paragraph{Definition of the group error.}
We first run dense attention on $\mathcal D_{\mathrm{rank}}$ and cache its
final vocabulary logits. Set every group's overall target to $0.9$
and evaluate the resulting sparse model against this dense reference.
Writing $\boldsymbol{\theta}^{\mathrm{ref}}=(0.9,\ldots,0.9)$, define
\begin{equation}
 d(\boldsymbol{\theta})=\frac{1}{|\mathcal I|}\sum_{(x,t)\in\mathcal I}
 d_{\mathrm{KL}}\!\left(
   \operatorname{softmax}(\mathbf{z}^d_{x,t})\,\middle\|\,
   \operatorname{softmax}(\mathbf{z}^s_{x,t}(\boldsymbol{\theta}))\right),
 \qquad d_{\mathrm{ref}}=d(\boldsymbol{\theta}^{\mathrm{ref}}),
 \label{eq:app_calibration_reference_kl}
\end{equation}
where $\mathcal I$ is the fixed set of example--position pairs drawn only
from $\mathcal D_{\mathrm{rank}}$.
The softmax is over the vocabulary, with the dense distribution as the
first argument of KL, consistent with Eq.~\ref{eq:predictive_goal}.
For each group $g$, independently reduce only its threshold by a factor
of $0.9$ and recompute the full model's final logits:
\begin{equation}
 \theta^{(g)}_h=
 \begin{cases}0.81,&h=g,\\0.9,&h\ne g,\end{cases}
 \qquad e_g:=d(\boldsymbol{\theta}^{(g)}),
 \qquad \Delta e_g:=e_g-d_{\mathrm{ref}}.
 \label{eq:app_calibration_group_error}
\end{equation}
All examples, prediction positions, and other settings stay fixed.
Each probe starts from $\boldsymbol{\theta}^{\mathrm{ref}}$, so reductions do not
accumulate across groups while building this ranking. We order groups by
increasing $e_g$; subtracting the common baseline $d_{\mathrm{ref}}$ gives
the identical ordering by $\Delta e_g$. Lower values indicate that the
single-group reduction better preserves the dense prediction distribution.
This KL error determines the trial order; the task score below determines
whether a trial reduction is accepted.

\paragraph{Search rule.}
Initialize every group at $\tau_0=0.9$ and evaluate its score on
$\mathcal D_{\mathrm{search}}$. Each trial changes only the current batch to
$\rho\tau_0$, with $\rho=0.9$, leaving previously accepted changes
in place. Accept a trial if its search-set score does not decrease.
If the score decreases, restore the accepted configuration and split
the batch into two contiguous halves in the error ordering. The lower-error
half is tested first. A rejected singleton retains $\tau_0$.
The procedure makes a single reduction decision per group; it does not
repeatedly multiply already accepted thresholds by $\rho$.
The search terminates when the stack is empty or when the predefined maximum
number of calibration steps $T_{\max}$ is reached. If the step limit is
reached first, groups in the remaining unvisited batches retain their current
thresholds.

\begin{algorithm}[!htbp]
 \caption{Error-Ranked Threshold Calibration}
 \label{alg:calibration}
 \small
 \begin{algorithmic}[1]
  \Statex \textbf{Input:} Groups $\mathcal G$; errors $e_g$ from
    Eq.~\ref{eq:app_calibration_group_error};
    initial threshold $\tau_0$; reduction factor $\rho=0.9$;
    fixed-search-set score $s(\boldsymbol{\theta})$;
    maximum calibration steps $T_{\max}$.
  \Statex \textbf{Output:} Frozen group thresholds $\boldsymbol{\theta}$.
  \State $\mathcal G\gets\operatorname{SortAscending}(\mathcal G,e_g)$
  \State $\boldsymbol{\theta}\gets(\tau_0,\ldots,\tau_0)$;
    $s_{\mathrm{best}}\gets s(\boldsymbol{\theta})$;
    $\mathcal Q\gets[\mathcal G]$ \Comment{LIFO stack}
  \State $t\gets0$
  \While{$\mathcal Q\ne\emptyset$ \textbf{and} $t<T_{\max}$}
   \State $\mathcal B\gets\operatorname{Pop}(\mathcal Q)$;
     $\boldsymbol{\theta}'\gets\boldsymbol{\theta}$
   \State $t\gets t+1$
   \State Set $\theta'_g\gets\rho\tau_0$ for every $g\in\mathcal B$
   \State $s'\gets s(\boldsymbol{\theta}')$
   \If{$s'\ge s_{\mathrm{best}}$}
    \State $\boldsymbol{\theta}\gets\boldsymbol{\theta}'$;
      $s_{\mathrm{best}}\gets s'$
   \ElsIf{$|\mathcal B|>1$}
    \State $(\mathcal B_{\mathrm{lo}},\mathcal B_{\mathrm{hi}})
      \gets\operatorname{Bisect}(\mathcal B)$
    \State $\operatorname{Push}(\mathcal Q,\mathcal B_{\mathrm{hi}})$;
      $\operatorname{Push}(\mathcal Q,\mathcal B_{\mathrm{lo}})$
   \EndIf
  \EndWhile
  \State \Return $\boldsymbol{\theta}$
 \end{algorithmic}
\end{algorithm}

\paragraph{Final configuration and calibration cost.}
All reported runs use $p_1=0.94$, $\tau_0=0.90$, and $\rho=0.90$.
Consequently, an unreduced group uses
$p_{2,g}=0.90/0.94\approx0.95745$, while an accepted reduced group uses
$p_{2,g}=0.81/0.94\approx0.86170$. The frozen artifact therefore needs only
one bit per group in addition to the shared constants: unreduced or reduced.
Its practical
effect is assessed by the module ablation in
Table~\ref{tab:module_ablation_128k}.

\end{document}